\documentclass[9pt,twocolumn]{extarticle}
\usepackage{authblk}
\usepackage[resetlabels,labeled]{multibib}
\newcites{S}{References}

\usepackage[utf8]{inputenc}
\usepackage[english]{babel}
\usepackage{amsmath,amsfonts,amssymb}
\usepackage{mathpazo} 
\usepackage[scaled]{helvet}
\usepackage[T1]{fontenc}
\usepackage{url}
\usepackage[colorlinks=true, allcolors=blue]{hyperref}

\usepackage{graphicx,xcolor}
\usepackage{colortbl}
\usepackage{booktabs}
\usepackage{algorithm}
\usepackage[noend]{algpseudocode}
\usepackage{changepage}
\usepackage[left=48pt,%
                right=42pt,%
                top=46pt,%
                bottom=60pt,%
                headheight=15pt,%
                headsep=10pt,%
                letterpaper,twoside]{geometry}%
\usepackage[labelfont={bf,sf},%
                labelsep=period,%
                figurename=Fig.,%
                singlelinecheck=off,%
                justification=RaggedRight]{caption}
\usepackage{amsmath,amssymb,graphicx}

\usepackage{physics}
\DeclareMathAlphabet\mathbfcal{OMS}{cmsy}{b}{n}
\newcommand{\bettershortstack}[2][c]{%
  \begin{tabular}[b]{@{}#1@{}}#2\end{tabular}%
}

\newcommand{\beginsupplement}{%
        \setcounter{table}{0}
        \renewcommand{\thetable}{S\arabic{table}}%
        \setcounter{figure}{0}
        \renewcommand{\thefigure}{S\arabic{figure}}%
     }

\title{Delocalized SPM rogue waves in normal dispersion cascaded supercontinuum generation}

\author[1,*]{Rasmus Eilk{\oe}r Hansen}
\author[1,3]{Rasmus Dybbro Engelsholm}
\author[1,2]{Christian Rosenberg Petersen}
\author[1,2,3]{Ole Bang}

\affil[1]{DTU Fotonik, Department of Photonics Engineering, Technical University of Denmark, 2800 Kgs. Lyngby, Denmark}
\affil[2]{NORBLIS IVS, Virumgade 35D, 2830 Virum, Denmark.}
\affil[3]{NKT Photonics A/S, Blokken 84, 3460 Birkerød, Denmark}

\affil[*]{Corresponding author: raeha@fotonik.dtu.dk}

\begin{document}

\maketitle

\begin{abstract}
In the numerical modelling of cascaded mid-infrared (IR) supercontinuum generation (SCG) we have studied how an ensemble of spectrally and temporally distributed solitons from the long-wavelength part of an SC evolves and interacts when coupled into the normal dispersion regime of a highly nonlinear chalcogenide fiber. This has revealed a novel fundamental phenomenon – the generation of a temporally and spectrally delocalized high energy rogue wave in the normal dispersion regime in the form of a strongly self-phase-modulation (SPM) broadened pulse. Along the local SPM shape the rogue wave is localized both temporally and spectrally. We demonstrate that this novel form of rogue wave is generated by inter-pulse Raman amplification between the SPM lobes of the many pulses causing the initially most delayed pulse to swallow the energy of all the other pulses. We further demonstrate that this novel type of rogue wave generation is a key effect in efficient long-wavelength mid-IR SCG based on the cascading of SC spectra and demonstrate how the mid-IR SC spectrum can be shaped by manipulating the rogue wave.
\end{abstract}
\section{Introduction}
\begin{figure}[htbp]
    \centering
    \fbox{\includegraphics[width = 0.96 \linewidth]{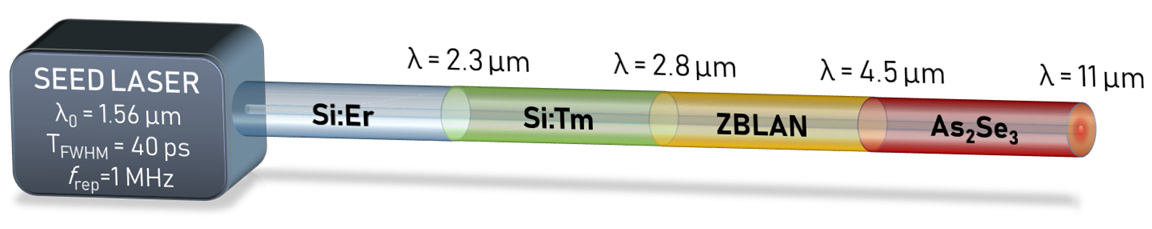}}
    \caption{Illustration of the full fiber cascade. The numbers above each fiber output is the long wavelength edge of the supercontinuum. The focus of this article is on the spectral evolution in the ZBLAN and $\mathrm{As_2Se_3}$ fibers.}
    \label{fig:Cascade}
\end{figure}

Ever since the discovery of temporally and/or spatially localized solitons and their particle-like behavior \cite{scott2003nonlinear}, their interaction has been the subject of intense research in physics and applied mathematics \cite{Interaction}. The outcome of soliton collisions has been shown to be highly dependent on the relative phase and amplitude of the two solitons, where in-phase solitons attract each other and out-of-phase solitons repel each other \cite{Mitschke}. In physical systems described by an integrable model, solitons have the beautiful property that they are able to collide and pass through each other without changing their shape or energy \cite{Integrable_Collision}, an effect that is often used as the definition of integrability of a nonlinear dynamical model, such as the nonlinear Schrödinger (NLS) equation describing optical fibers and a vast range of other nonlinear optical systems \cite{Zakharov}. Non-integrable perturbations from effects, such as higher-order dispersion or Raman scattering, will drastically change the soliton collision properties by allowing energy transfer between solitons \cite{Chi}. This might not seem so drastical at first if one considers only one collision between two solitons, but if one considers the outcome of a series of collisions between a large number of different solitons under the influence of a perturbation that leads to an on average preferential energy transfer, e.g., from low amplitude to high amplitude solitons, then the outcome can be an extremely localized soliton with an extremely high amplitude, also known as a rogue wave \cite{Islam:89,Bang}. \\
Rogue waves appear in two fundamentally different forms: One type appears out-of-nowhere and disappears just as fast, such as the both spatially and temporally localized Peregrine soliton solution of the NLS equation, which locally at the point of maximum compression has an intensity 9 times higher than the background \cite{Kibler}. Here we focus on the second type of rogue wave that is the result of many collisions \cite{Islam:89}, which has a relatively long life time, making them also biologically relevant in for example DNA denaturation \cite{Bang}.  \\
One particularly useful testing gronud for the study and observation of rogue waves has for a long time been supercontinuum generation (SCG) in optical fibers pumped in the anomalous dispersion regime with long pulses. The first work in 1989 of Islam et al. \cite{Islam:89} (here termed narrow, high-intensity solitons) dates back to before the work in 1993-95 on biological rogue waves by Peyrard et al. \cite{Dauxois,Bang} (here termed large amplitude breathers or high-energy localized vibrational modes). In this case SCG is initiated by modulational instability (MI) growing from noise in frequency bands symmetrically displaced by $\Omega_m \approx (2 \gamma P_0/\abs{\beta_2})
^{1/2}$ from the pump frequency and breaking up the pump pulse into a number of solitons over an MI gain length of $1/(\gamma P_0)$, where $P_0$, is the pump peak power, $\gamma$ is the fiber nonlinearity, and $\beta_2$ is the group-velocity dispersion \cite{agrawal2013nonlinear}. The pulse duration of the generated solitons is roughly $\pi/\Omega_m$, and thus both the pulse length of the solitons and the distance over which they are generated depend on the pump peak power. The result will be a distributed spectrum of different solitons, since the high peak power in the center generates short solitons early in the fiber, while the low peak power wings will generate longer solitons after a longer propagation distance, as demonstrated by Islam et al. \cite{Islam:89}. \\
The rather controlled generation of a distributed soliton spectrum though MI is very important for the generation of rogue waves. The other important mechanism is Raman scattering, which represents a perturbation to the integrable NLS equation that leads to two key effects: (1) intra-pulse stimulated Raman scattering (SRS) casusing a continuous red-shift of a soliton, also known as the soliton self-frequency shift (SSFS), which scales inversely with pulse duration \cite{Gordon} and (2) a preferential transfer of energy from low- to high-amplitude solitons in a collision, mediated by the phase-insensitive inter-pulse SRS as demonstrated by Luan et al. in 2006 \cite{Luan}. Since solitons of different pulse durations red-shift by different rates due to SSFS, a distributed spectrum will inevitably lead to collisions of solitons that are different. For increasing pulse energy SCG will involve the generation of an increasing number of solitons, and thus the number of collisions will also increase. The Kerr effect will lead to an energy transfer that depends sensitively on the relative phase and amplitude of the colliding solitons, but inter-pulse SRS adds to this with a preferential transfer of energy from the most blue-shifted to the most red-shifted soliton, which is typically also the largest due to SSFS \cite{Luan}. This means that potentially a rare event could occur, in which the conditions are just right for one soliton to gain energy from many collisions and thereby obtain a very high amplitude and narrow width. Due to the Raman-induced SSFS in optical fibers, these high-amplitude short solitons would red-shift strongly and inevitably be located in the most red part of the SC spectrum. This was already observed numerically by Islam et al. in 1989, who also experimentally proved that different red parts of the spectrum were indeed not present in all pulses \cite{Islam:89}. In 2006 Frosz et al. numerically investigated CW-pumped SCG and demonstrated also the generation of a very short and high-amplitude soliton with a huge red-shift \cite{Frosz}. This set the stage for the work of Solli et al., who in 2007 measured the L-shaped statistics of these rare high-energy waves for the first time and gave them the name optical rogue waves, in analogy with oceanic rogue waves \cite{Solli}. It was later demonstrated that in fact the Raman effect is not necessary to generate optical rogue waves, even just 3rd order dispersion provides a sufficient non-integrable perturbation to provide on-average preferential energy transfer in soliton collisions and generate rogue waves \cite{Genty}. Since then the field of optical rogue waves has become an important and very rich scientific field with strong parallels to oceanography and hydrodynamics because of the common dynamical model – the NLS equation \cite{review}. \\
So far, fiber-optical rogue waves have only been demonstrated as highly localized high peak power solitons existing in the anomalous dispersion region because of the self-focusing nonlinearity of the fibers. In other words, they are strongly linked to the initial generation of a distributed spectrum of localized solitons, which requires either MI (of long pump pulses) or soliton fission (of short pump pulses). In fact also the spatially (along the direction of propagation) and temporally localized Peregrine rogue wave requires anomalous dispersion \cite{Kibler}. \\
\textit{Here we demonstrate the first high-energy rogue wave, which is generated in the normal dispersion region, where no solitons exist, and which is both temporally and spectrally delocalized. This normal dispersion rogue wave takes the shape of a very high energy self-phase modulation (SPM) wave.} \\
We observe the generation of the novel SPM rogue wave in the numerical modelling of cascaded mid-infrared (mid-IR) SCG in which a large ensemble of spectrally and temporally distributed solitons from the long-wavelength part of an SC generated in one fiber (here a ZBLAN fiber) is coupled into the normal dispersion regime of another fiber (here a highly nonlinear chalcogenide fiber). We demonstrate that the SPM rogue wave is generated by inter-pulse Raman amplification between the SPM lobes of the many pulses, causing the initially most delayed pulse to gradually swallow the energy of all the other SPM broadened pulses as they begin to overlap temporally due to dispersion, while also being within the Raman gain band (extending to about 10 THz in the chalcogenide fiber). \\
Cascaded SCG is currently one of the most promising routes for a practical and high brightness light source covering the important mid-IR spectral region from $2-12 \mu \mathrm{m}$, with applications including: Chemical detection \cite{Chemical_Detection}, tissue microspectroscopy \cite{Tissue_Imaging,ChemicalMapping}, and optical coherence tomography \cite{OCT}. We consider the specific cascaded mid-IR SC laser shown in Fig.~\ref{fig:Cascade}.
and demonstrate that the SPM rogue wave generation is a key effect in efficient long-wavelength mid-IR SCG based on the cascading of SC spectra. Thereby, we demonstrate how the mid-IR SC spectrum can be shaped by manipulating the SPM rogue wave. In particular, we demonstrate that the SPM rogue wave and the left over parts of the other SPM lobes together act as a spectrally localized collective structure (here about 500-800 nm broad) whose center wavelength is slowly red-shifting towards the zero-dispersion wavelength (ZDW) through the inter-pulse Raman amplification between the SPM lobes, finally stopping at a certain distance before the ZDW where the dispersion is so weak that effectively no further temporal overlap is possible. This slowly red-shifting collective structure was recently reported in a Master's project \cite{RasmusMSC} and in a similar study of cascaded mid-IR SCG by coupling a ZBLAN fiber SC into a chalcogenide fiber with normal dispersion \cite{Venck}. In \cite{Venck} the authors explain the red-shift as being due to intra-pulse Raman scattering, but here our detailed investigations, based on series of spectrograms, clarifies that the slow red-shift is actually  due to collective inter-pulse Raman amplification and the generation of an SPM rogue wave.

\section{The numerical model and fiber parameters}

To model the fiber cascade shown in Fig.~\ref{fig:Cascade}
the propagation of the pulse envelope of a single scalar mode $G(t,z)$ (or $\Tilde{G}(\Omega,z)$ in the frequency domain) is simulated through the generalized NLS equation (GNLSE)
\begin{equation}
    \begin{split}
      & \frac{\partial}{\partial z}\left[\exp\left( i \beta_{\mathrm{eff}}(\Omega, \Omega_P, \Omega_0) z \right)^* \right] \Tilde{G}(\Omega,z) \\ & =
       i \gamma(\Omega) K(\Omega,\Omega_0) \exp\left( i \beta_{\mathrm{eff}}(\Omega, \Omega_P, \Omega_0) z \right)^* \label{eq:GNLSE} \\
       & \mathcal{F} \left\{ H \mathcal{F}^{-1} \left[ \Tilde{R}(\Omega) \mathcal{F} \Big[ H^* H \Big] \right]
       \right\}
    \end{split}
\end{equation}
as introduced in \cite{Dybbro_Noise}, where $\mathcal{F}\{...\}$ and $\mathcal{F}^{-1}\{...\}$ are the Fourier transform and its inverse, respectively, superscript $*$ is the complex conjugate, and 
\begin{equation}
    \begin{split}
    K(\Omega, \Omega_0) & = \left[ \frac{\mathrm{A_{eff}}(\Omega_0)}{\mathrm{A_{eff}}(\Omega)} \right]^{\frac{1}{4}}\\ 
    H &= H(t,z) = \mathcal{F}^{-1}\left[\Tilde{G}(\Omega,z) K(\Omega,\Omega_0) \right]
    \end{split}
\end{equation}
The introduction of $K(\Omega,\Omega_0)$ treats mode profile dispersion, as described in \cite{MFD_Jesper}.
The effective propagation constant $\beta_{\mathrm{eff}}=\beta_{\mathrm{eff}}(\Omega,\Omega_p,\Omega_0)$ is defined as
\begin{equation}
    \beta_{\mathrm{eff}}(\Omega,\Omega_p,\Omega_0) \equiv \beta(\Omega) - \beta_0(\Omega_0) - \beta_1(\Omega_p)[\Omega - \Omega_0] 
\end{equation}
where the propagation constant $\beta(\Omega)$ is related to the effective index by $\beta(\Omega) = \Omega n_{\mathrm{eff}}(\Omega)/c$.
As such the full frequency dependant propagation constant is used, where the evaluation of the inverse group velocity $\beta_1(\Omega_p)$ at the pump frequency, $\Omega_P$, ensures that the pump wavelength is stationary in the time domain. $\Omega$ is the physical frequency, and $\Omega_0$ is the center of the frequency grid. 
Thus, the pulse envelope $\tilde{G}(\Omega,z)$ is related to the real pulse envelope $\tilde{A}(\Omega,z)$, used by e.g Agrawal \cite{agrawal2013nonlinear}, by a simple phaseshift as $\tilde{G}(\Omega,z) = \exp(i \beta_{\mathrm{eff}} z) \tilde{A}(\Omega,z)$. \\
\begin{figure}[h]
    \centering
    \fbox{\includegraphics[width = 0.96 \linewidth]{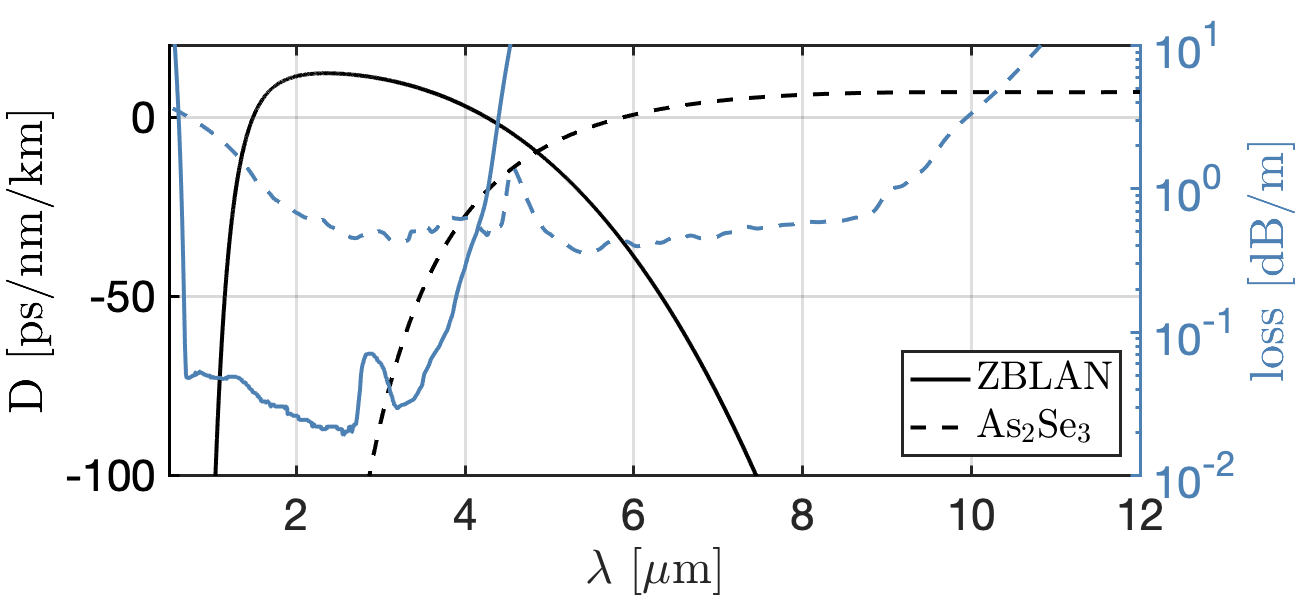}}
    \caption{Dispersion and loss profiles of the ZBLAN and $\mathrm{As_2Se_3}$ fibers. The dispersion of the ZBLAN fiber was reported in \cite{ZBLAN_Dispersion}. The dispersion of the $\mathrm{As_2Se_3}$ fiber has been obtained in COMSOL with the core index given in \cite{As2SeMaterialIndex}. The loss of the ZBLAN fiber was measured in-house and the loss of the $\mathrm{As_2Se_3}$ fiber was measured by the manufacturer IRflex, and is given in \cite{Selenide_loss}.}
    \label{fig:Disp_N_Loss}
\end{figure}
In Fig. \ref{fig:Disp_N_Loss} the attenuation and dispersion $D = - \frac{\lambda}{c} \pdv[2]{n_{\mathrm{eff}}}{\lambda}$ of the ZBLAN and $\mathrm{As_2Se_3}$ fibers are shown. The ZBLAN fiber has two ZDWs at approximately $1.5 \mu \mathrm{m}$ and $4.2 \mu \mathrm{m}$, with an anomalous dispersion regime in between them, allowing soliton propagation. The $\mathrm{As_2Se_3}$ fiber has only one ZDW at approximately $6 \mu \mathrm{m}$, which means that the output of the ZBLAN fiber will be coupled into the normal dispersion regime. The background loss of the $\mathrm{As_2Se_3}$ fiber is orders of magnitude higher than that of the ZBLAN fiber, guidance is however allowed much further into the mid-IR, and the nonlinearity is much higher.
The frequency dependent nonlinear parameter $\gamma(\Omega)$ is given by
\begin{equation}
    \gamma(\Omega) = \frac{n_2 n_0^2 \Omega}{c n_{eff}(\Omega)^2 A_{\mathrm{eff}}(\Omega_0)}
\end{equation}
where $n_0$ is the refractive index at $\Omega_0$, and $n_2$ is the nonlinear refractive index. The usual definition of the effective area is used:
\begin{equation}
    A_{\mathrm{eff}}(\Omega) = \frac{\qty(\int \int  \abs{\mathbfcal{E}(\Omega,x,y)}^2\mathrm{d}x \mathrm{d} y)^{2}}{\int \int 
    \abs{\mathbfcal{E}(\Omega,x,y)}^4 \mathrm{d}x\mathrm{d}y}
\end{equation}
In this implementation the transverse fields $\mathbfcal{E}(\Omega,x,y)$ have been normalized such that the unit of the envelope in the time domain is $\sqrt{\mathrm{W}}$.
Finally, the nonlinear response function of the material is given by
\begin{equation}
    \Tilde{R}(\Omega) = 1-f_\mathrm{R} + f_\mathrm{R}\Tilde{h}_r(\Omega)
\end{equation}
where the term $1-f_{\rm R}$ is the instantaneous Kerr response, $\Tilde{h}_r(\Omega)$ is the delayed Raman response function, and $f_R$ is the fractional Raman contribution. The Raman reponse functions are shown in Fig. \ref{fig:Raman} and the key fiber parameters are given in Table \ref{tab:fiber_Param}.\\
\begin{figure}[htbp]
    \centering
    \fbox{\includegraphics[width = 0.96 \linewidth]{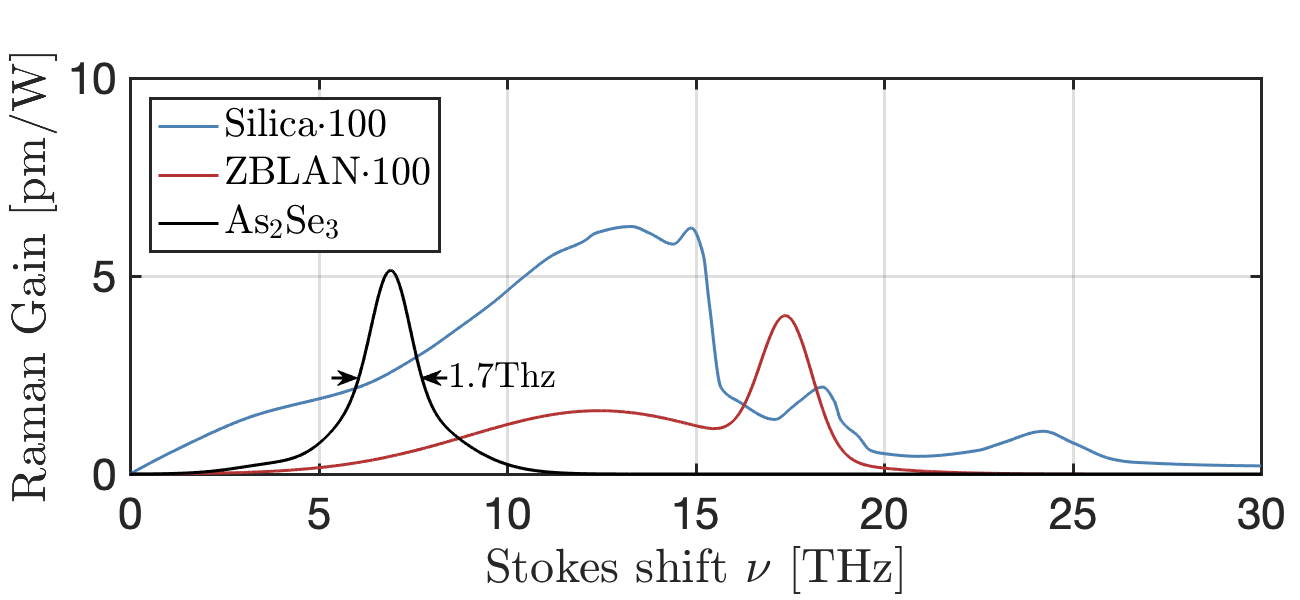}}
    \caption{The Raman gain for silica, ZBLAN, and $\mathrm{As_2Se_3}$. The silica gain has been obtained from \cite{agrawal2013nonlinear}, ZBLAN from \cite{ZBLAN_Raman}, and $\mathrm{As_2Se_3}$ from \cite{As2Se3_Raman}. Notice that the silica and ZBLAN gain are multiplied by 100.}
    \label{fig:Raman}
\end{figure} \\
\begin{table}[h]
    \centering
    \begin{tabular}{c c c c c c} \hline
     \bettershortstack{fiber \\ \vphantom{$\mu$}} & \bettershortstack{ a \\ $[\mu \mathrm{m}]$} & \bettershortstack{NA \\ \vphantom{$\mu$}} & \bettershortstack{$n_2 \cdot 10^{20}$ \\ $[\mathrm{m^2/W}]$} &
     \bettershortstack{$f_R$ \\  \vphantom{$\mu$}} & \bettershortstack{Manufacturer \\ \vphantom{$\mu$}} \\  \hline
    ZBLAN     & 3.5 & 0.265 & 2.1 & 0.0969 & FiberLabs\\
    $\mathrm{As_2Se_3}$     & 6 & 0.76 & 1500 & 0.0103 & IRflex  \\ \hline
    \end{tabular}
    \caption{Constant fiber parameters in the simulations, where a is the fiber core radius and NA is the numerical aperture.}
    \label{tab:fiber_Param}
\end{table}
In all our simulations we use $N_t = 2^{19}$ grid points, a temporal resolution of $\mathrm{d}t = 1.5 \mathrm{fs}$, and a central angular frequency of $\Omega_0 = 2 \pi \cdot  350 \cdot  10^{12} \mathrm{rad \cdot s}^{-1}$. The narrowest solitons observed are approximately 20 fs, which means the temporal resolution should be adequate. This gives a wavelength range that spans from $438.7 \mathrm{nm}$ to $17989 \mathrm{nm}$. We use a variable step-size controller, based on an embedded fourth and fifth order Runge-Kutta stepper. Due to the high nonlinear refractive index of $\mathrm{As_2Se_3}$, proper convergence necessitates step-sizes as small as $0.1 \mu \mathrm{m}$. \\
In all the presented figures illustrating the evolution of the power spectral density (PSD) as a function of propagation distance $z$ a running spectral average has been applied with a stepsize of $3 \mathrm{nm}$, and a bandwidth of $15  \mathrm{nm}$. All spectrograms are calculated using a Hamming window with a width of $150 \mathrm{fs}$.

\section{Cascaded supercontinuum generation}

\begin{figure}[htbp]
    \centering
    \fbox{\includegraphics[width =  0.96\linewidth]{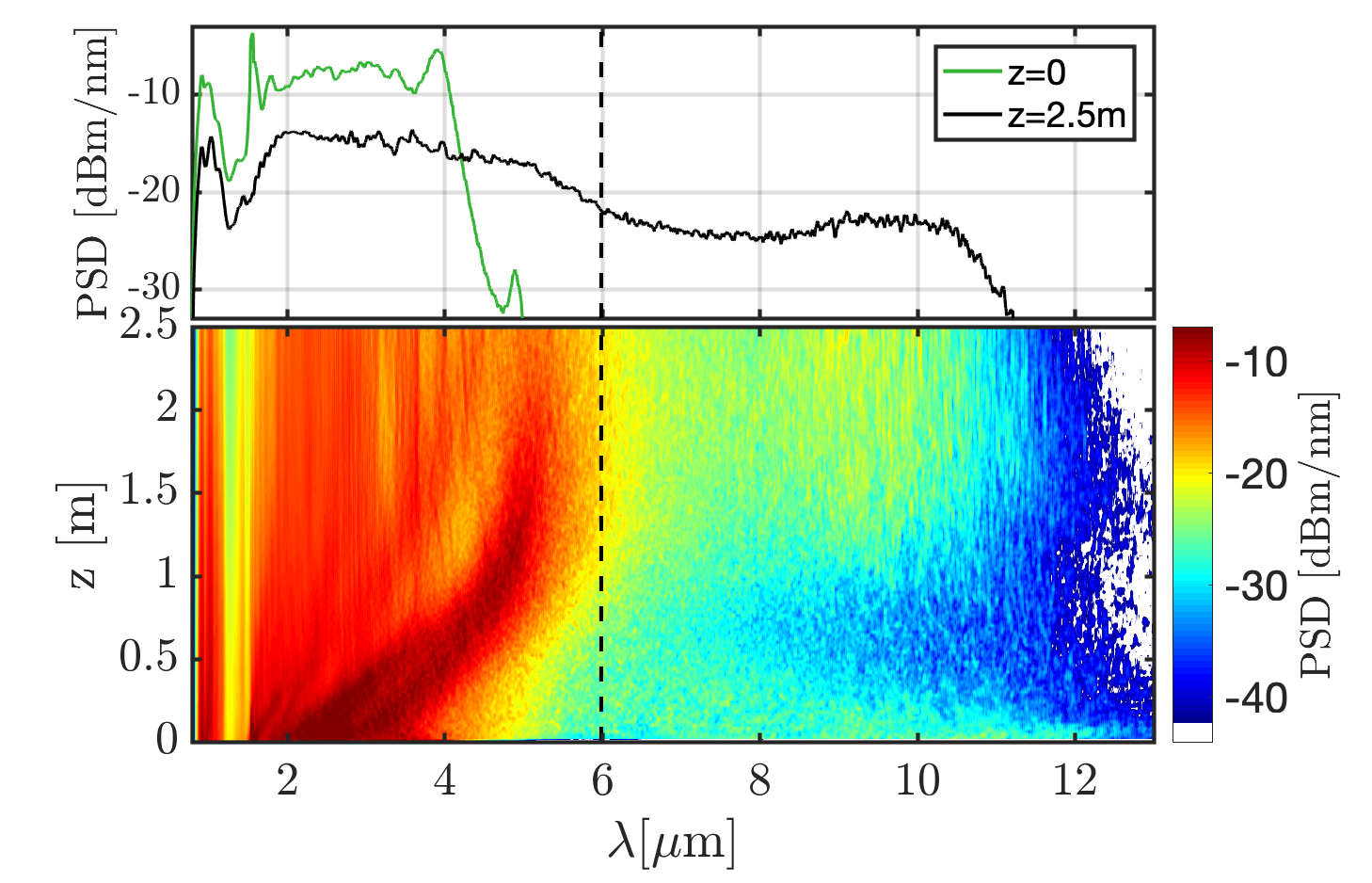}}
    \caption{Bottom: Spectral evolution in the $\mathrm{As_2Se_3}$ fiber of the full ZBLAN fiber output of a single noise seed. Top: Input and output spectra taken as an average of 10 noise seeds.}
    \label{fig:Full_Evo}
\end{figure}

\begin{figure}[h]
    \centering
    \fbox{\includegraphics[width = 0.96 \linewidth]{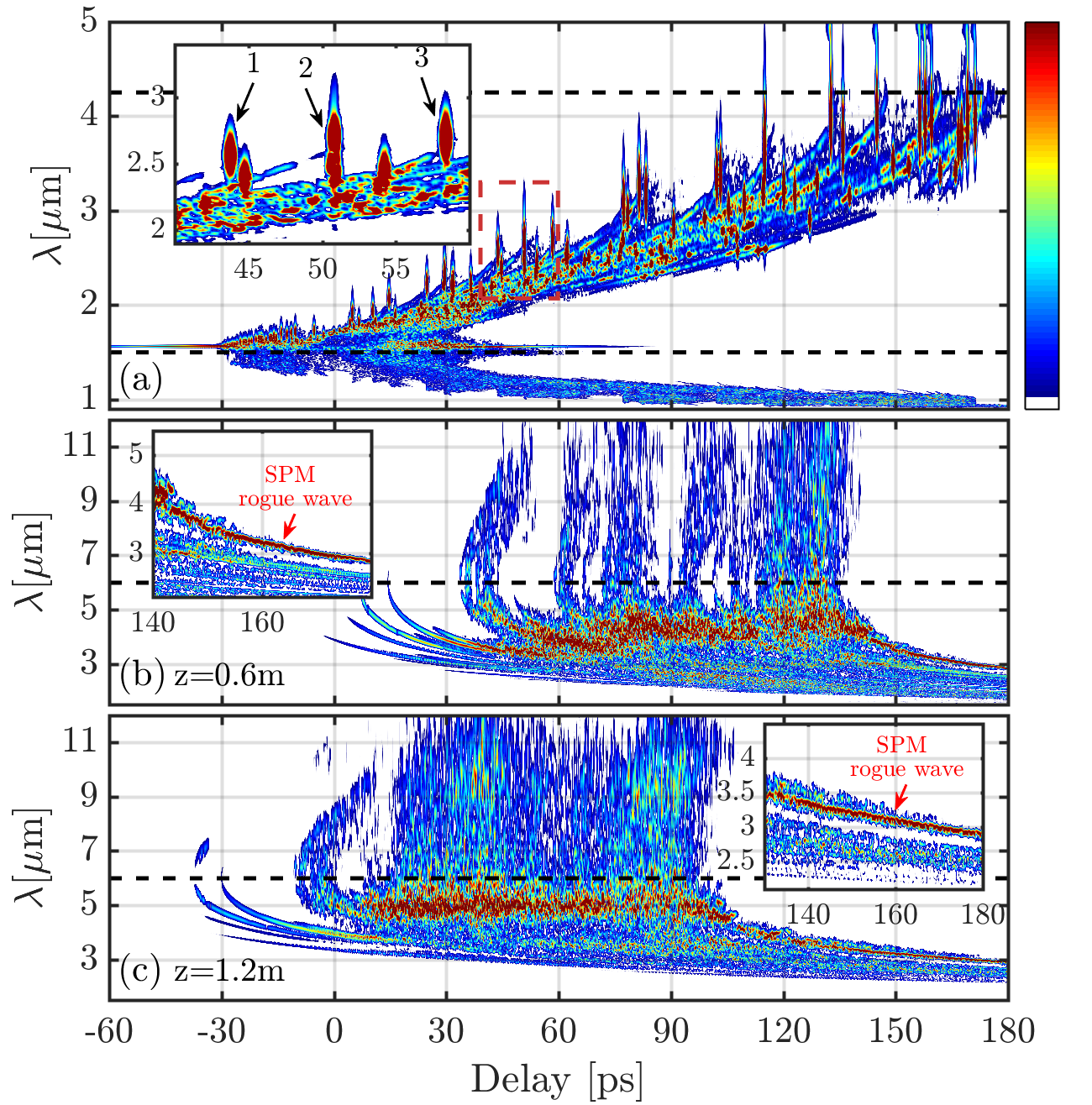}}
    \caption{(a) Spectrogram of the ZBLAN fiber output with a zoom on the red dashed region with solitons 1-3. The horizontal dashed lines indicate the ZDWs of the ZBLAN fiber with anomalous dispersion in between them.
    (b-c) show spectrograms after 0.6m (b) and 1.2m (c) of propagation in the $\mathrm{As_2Se_3}$ fiber with zooms on the SPM rogue wave. The limits to the color axis in all plots (insets) is between the minimum (10\% of the maximum) and maximum at $z=0$.}
    \label{fig:ZBLAN_output}
\end{figure}

In the following we focus on the $\mathrm{As_2Se_3}$ fiber stage of the cascaded SC source shown in Fig. \ref{fig:Cascade}. The input to the $\mathrm{As_2Se_3}$ fiber is the output of the ZBLAN fiber, which is obtained by propagating noise seeded Gaussian shaped 40ps pulses from a directly modulated 1560nm seed diode through a 5m long Er-doped silica fiber amplifier (EDFA), followed by a 1.5m long Tm-doped fiber amplifier (TDFA), similar to the one described in \cite{ZBLAN_Dispersion}, and finally the 7m long ZBLAN fiber. During amplification in the EDFA, the input pulse undergoes MI and breaks up into a number of solitons to generate an in-amplifier SC \cite{In_Amp_SCG, Kyei}, resulting in an SC with a spectral edge around 2.3$\mu \mathrm{m}$ and 800mW of average power at $1\mathrm{MHz}$ repetition rate. In the TDFA the spectral edge is extended to 2.8$\mu$m before being coupled into the 7m ZBLAN fiber to generate the ensemble averaged output spectrum shown in the top of Fig. \ref{fig:Full_Evo} (averaged over 10 noise seeds), which reaches 4.5$\mu$m and has 220mW average power at 1MHz repetition rate. Essentially, we now consider how 10 different versions of the single shot (one noise seed) ZBLAN output spectrogram shown in Fig. \ref{fig:ZBLAN_output}(a) evolve in the $\mathrm{As_2Se_3}$ fiber. All detailed parameters of all fibers and amplifiers not given here can be found in the supplementary material (see also \cite{RasmusMSC}). \\
The result of propagating the full output of the ZBLAN fiber in 2.5m $\mathrm{As_2Se_3}$ fiber is shown in Fig. \ref{fig:Full_Evo}, with the input and output spectra shown above. The spectrum was averaged over an ensemble of 10 noise seeds, with both one-photon-per-mode noise and 1 \% relative intensity noise. A part of the power is seen to immediately shift across the ZDW at 6$\mu$m out to about 12$\mu$m, but as the pulse propagates along the fiber, the mid-IR power above 6$\mu$m attenuates rather quickly again. However, an about 500-800nm broad localized excitation carrying a major part of the power is seen to slowly red-shift from $\sim 3 \mu \mathrm{m}$ to $\sim 5 \mu \mathrm{m}$, while also pushing power across the ZDW and far into the mid-IR. This red-shifting collective excitation was also recently observed in \cite{RasmusMSC,Venck}. The main peak in the spectrum ends up approximately at $5 \mu \mathrm{m}$ after 1.2m of propagation. As a result the mid-IR power above 6$\mu$m peaks after 2m of propagation. At the output, the spectrum reaches 11$\mu \mathrm{m}$ and at 1MHz repetition rate the total power is 77mW with 12mW above the ZDW at 6$\mu \mathrm{m}$.\\
Understanding the physics behind the slowly red-shifting high PSD localized structure is the focus of this article. Using detailed spectrograms we will show that it is due to a collective effect of inter-pulse Raman amplification between a large number of pulses undergoing SPM, which finally leads to the generation of a novel type of high-energy SPM rogue wave.  \\
In Fig. \ref{fig:ZBLAN_output} we show spectrograms of the power distribution $\abs{G(z,t)}^2=|A(z,t)|^2$ of the ZBLAN fiber output and how it looks after having propagated 0.6m and 1.2m in the $\mathrm{As_2Se_3}$ fiber.
At the ZBLAN fiber output, seen in Fig. \ref{fig:ZBLAN_output}(a), a large number of solitons have been created in the anomalous dispersion regime between the ZDWs at 1.5$\mu$m and 4.2$\mu$m (dashed black lines), while in the normal dispersion regimes dispersive waves have been generated by the solitons, giving a spectrum extending from $1 \mu \mathrm{m}$ to about $4.5 \mu \mathrm{m}$. The inset shows a close up of the area marked by the red dashed rectangle, in which several intense and distinct solitons are clearly visible. In the following section the three numbered solitons 1-3 will be investigated separately to clearly illustrate the fundamental physics behind the observed spectral evolution. The pulse parameters of the three solitons (amplitude fitted to a sech profile) are given in Table \ref{tab:PulseParameters}.
\begin{table}[h]
    \centering
    \begin{tabular}{c c c c c c c} \hline
        \# & \bettershortstack{$P_0$ \\ $[\mathrm{kW}]$} & \bettershortstack{$\mathrm{T_{FWHM}}$ \\ $[\mathrm{fs}]$} &
        \bettershortstack{$\lambda_0$ \\ $[\mu \mathrm{m}]$} &
        \bettershortstack{$\gamma$ \\ $[\mathrm{1/(Wm)}]$} &
        \bettershortstack{$\beta_2$ \\ $[\mathrm{ps^2/m}]$} &
        \bettershortstack{$z_{\mathrm{OWB}}$ \\ $[\mathrm{mm}]$} \\ \hline 
        1 & 42.8 & 50 & 2.55 & 0.58 & 523 & 1.3\\ 
        2 & 78.6 & 30 & 2.72 & 0.54 & 476 & 0.6\\ 
        3 & 44.2 & 90 & 2.68 & 0.55 & 487 & 2.5\\ \hline
    \end{tabular}
    \caption{Parameters for solitons 1-3 marked in Fig. \ref{fig:ZBLAN_output}(a): Peak power $P_0$, temporal Full Width Half Max $T_{\rm FWHM}$ (measured in power), center wavelength $\lambda_0$, group velocity dispersion $\beta_2$, fiber nonlinearity $\gamma$, and OWB distance $z_{\mathrm{OWB}}$ calculated from the analytical formula given in \cite{Anderson_Wave_Breaking}, all evaluated at $\lambda_0$.}
    \label{tab:PulseParameters}
\end{table}
After 0.6m a considerable amount of power has been moved above the ZDW in the $\mathrm{As_2Se_3}$ fiber, but the bulk of the energy is now concentrated in a narrow band around $4 \mu \mathrm{m}$ as also clearly visible in Fig. \ref{fig:Full_Evo}. Several temporally elongated SPM profiles are visible and from the inset it is clear that the most delayed SPM profile has much higher PSD than the others. After 1.2m the localized band has red-shifted to an even narrower band closer to the ZDW at approximately $5 \mu \mathrm{m}$, as also seen in Fig. \ref{fig:Full_Evo}. Less SPM profiles are now visible and the most delayed is still the one with the highest PSD. A new type of rogue wave with an SPM shape appears to have been formed To further investigate this behaviour, we now focus on the interaction of the three solitons marked in Fig. \ref{fig:ZBLAN_output}(a).
\section{Inter-pulse Raman amplification and the SPM rogue wave}
\begin{figure}[!ht]
    \centering
    \fbox{\includegraphics[width = 0.96 \linewidth]{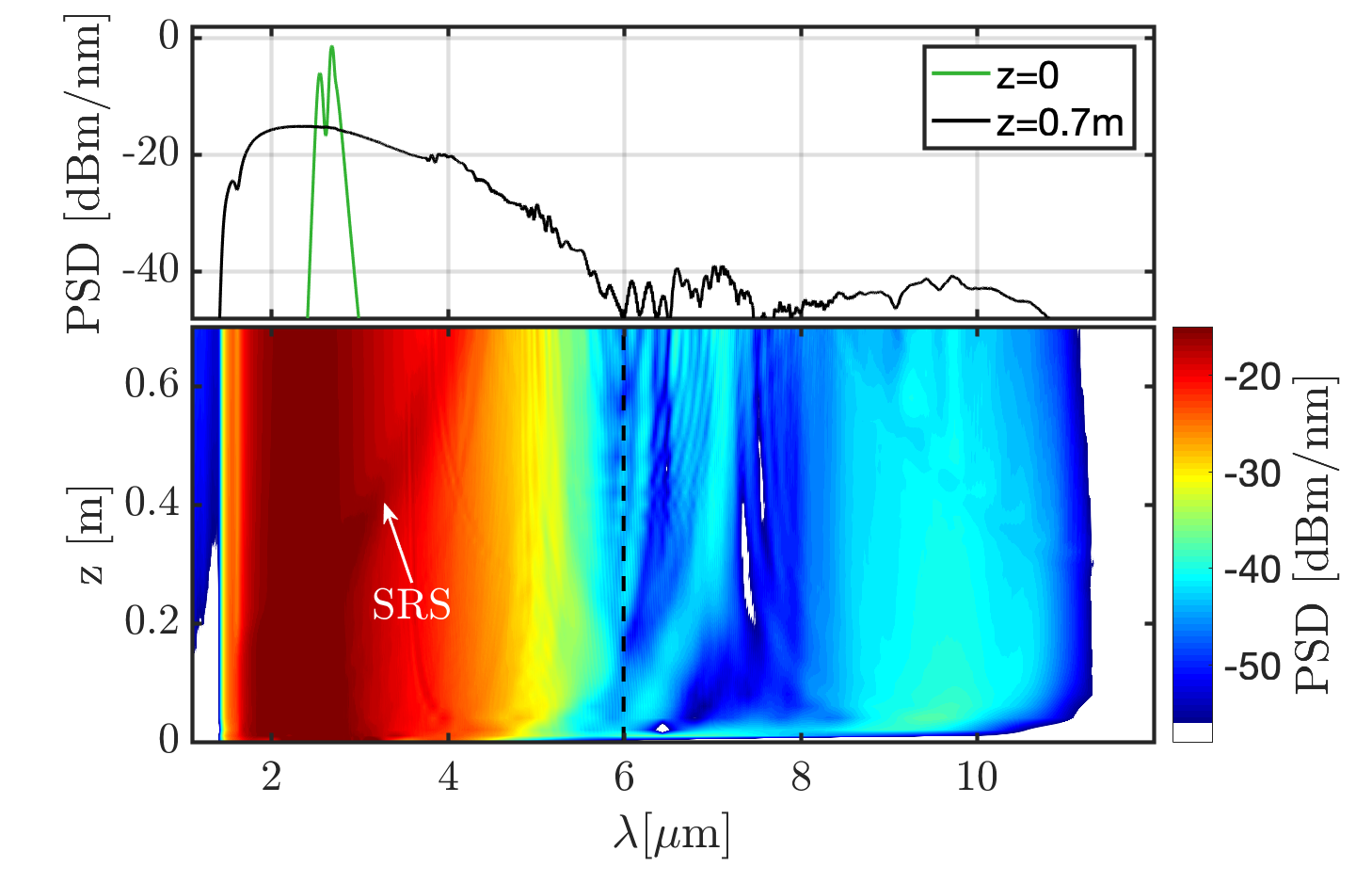}}
    \caption{Spectral evolution of solitons 1-3 to z=0.7m with input and output spectra. The evolving localized structure is marked by SRS.}
    \label{fig:3Solitons}
\end{figure}
\begin{figure}[!ht]
    \centering
    \fbox{\includegraphics[width = 0.97 \linewidth]{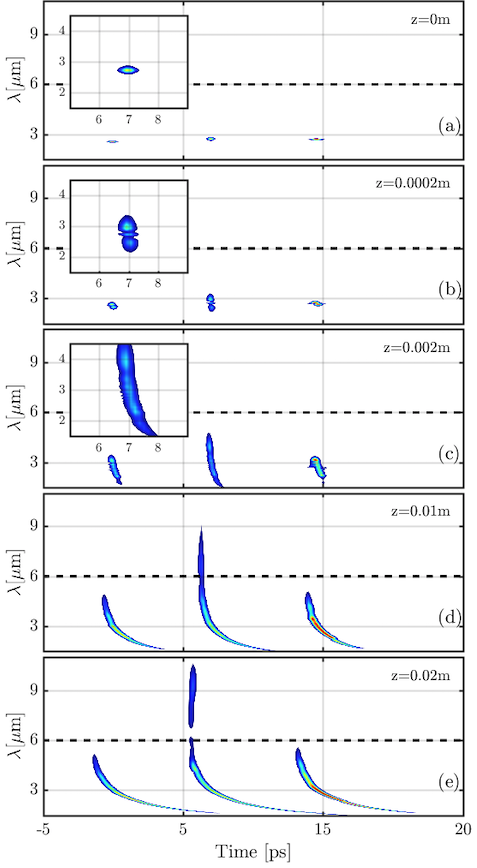}}
    \caption{Spectrograms showing the initial evolution of solitons 1-3 at selected distances out to z=2cm in the $\mathrm{As_2Se_3}$ fiber. Dashed lines indicate the ZDW and insets show a magnified image of soliton 2. The colors are not to scale between the images, but have been chosen to give clear individual contrast.}
    \label{fig:3Sol_Spectrogram}
\end{figure}
\begin{figure}[!ht]
    \centering
    \fbox{\includegraphics[width = 0.97 \linewidth]{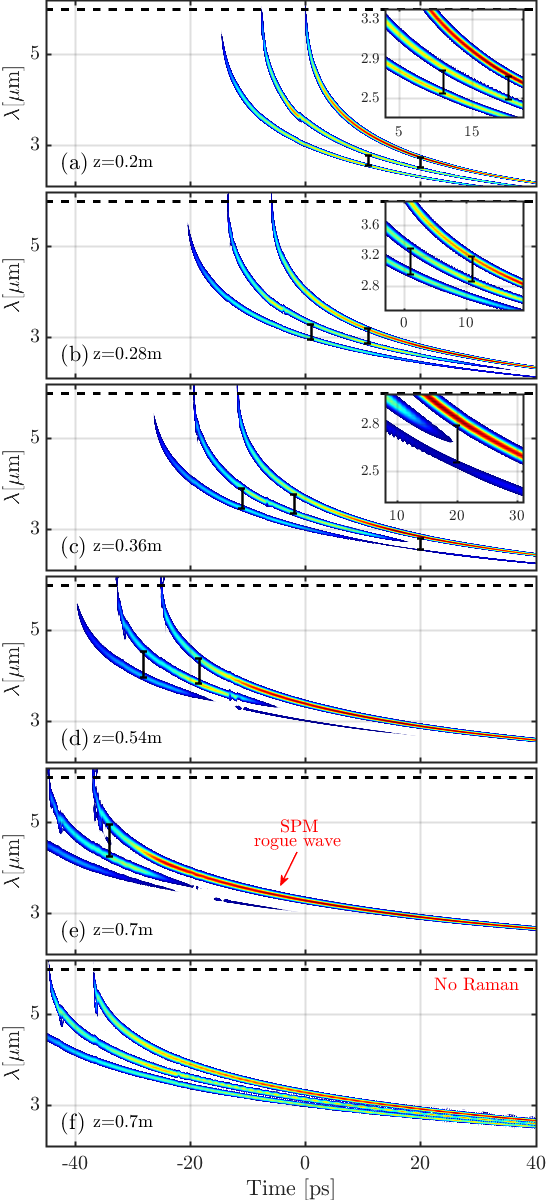}}
    \caption{Spectrograms at selected distances between z=20cm and 60cm in the $\mathrm{As_2Se_3}$ fiber, where the SPM pulses start to interact and generate an SPM rogue wave. The dashed lines indicate the ZDW. The colors are not to scale between the images, but have been chosen to give clear individual contrast. The small black indicators mark the 10THz Raman gain bandwidth. The spectrograms (e) and (f) are the same, except that Raman interaction was turned off in (f), by setting $f_R = 0$. The lower limit on the coloraxis in (f) is set to 10\% of max, to illustrate the individual pulse tails more clearly.}
    \label{fig:3Sol_Spectrogram_Last}
\end{figure}

The spectral evolution of only the three solitons selected from the ZBLAN output spectrum [see Fig. \ref{fig:ZBLAN_output}(a)] is shown in Fig. \ref{fig:3Solitons}. From now on we refer to them as pulses when discussing their evolution inside the $\mathrm{As_2Se_3}$ fiber. 
Initially, very strong spectral broadening due to SPM happens almost immediately due to the extremely high nonlinearity of the fiber, which shifts energy across the ZDW at 6$\mu \mathrm{m}$, just as in the evolution of the full spectrum with hundreds of solitons in Fig. \ref{fig:Full_Evo}. Even with only these 3 pulses starting relatively far apart a localized high PSD structure again becomes visible at around 0.3m and continuously red-shifts towards the ZDW. The localized structure is marked SRS, but at this point we have not shown whether it is inter-pulse or intra-pulse SRS that generates it. The fact that the structure appears later than for the full spectrum with many more temporally much closer solitons seems to indicate that it is an inter-pulse interaction effect. \\
To look even closer into the dynamics we use spectrograms. In Fig. \ref{fig:3Sol_Spectrogram} we focus on the initial dynamics out to z=2cm, where the 3 pulses evolve individually without significant overlap.
In Fig. \ref{fig:3Sol_Spectrogram}(b) we see the clear signature of initial SPM for all pulses at z=0.2mm. As expected SPM happens faster for pulse 2 because it has the highest peak power (see Table \ref{tab:PulseParameters}). At z=2mm dispersion has started to shift the SPM sidelobes temporally, and optical wave breaking (OWB) has started to set in for pulse 1 and 2, initially at the tails where the dispersion is the highest. At z=1cm pulse 2 has spectrally broadened to the ZDW and started to push energy across the ZDW, while pulse 3 now also clearly is undergoing OWB. At z=2cm pulse 2 has shed off what appears as a soliton in the anomalous dispersion regime above the ZDW and left is now in the normal dispersion regime three elongated SPM pulses still without overlap. \\
In Fig. \ref{fig:3Sol_Spectrogram_Last} we focus on the normal dispersion region and the evolution between z=20cm and z=60cm, in which the SPM pulses begin to overlap enough due to dispersion to start interacting through SRS. The interaction will be through inter-pulse Raman amplification, which will be effective for frequency separations up to approximately 10THz and strongest at about 7THz, as shown in Fig. \ref{fig:Raman}. In the spectrograms we mark the 10THz separation with a black indicator at the wavelength where it first connects two SPM lobes to show when the SPM lobes can transfer power between each other and when not. \\
In Fig. \ref{fig:3Sol_Spectrogram_Last}(a) for z=20cm we see that the tail of pulse 2 now can transfer energy to the most delayed pulse 3 at wavelength below approximately 2.8$\mu$m, while energy can be transferred from pulse 1 to 2 at wavelengths below 2.7$\mu$m. This is illustrated by the 10THz Raman indicator on pulse 2 (pulse 1) exactly reaching pulse 3 (pulse 2) at 2.8$\mu \mathrm{m}$ (2.7$\mu \mathrm{m}$).
The energy transfer between the SPM lobes is clearly visible in Fig. \ref{fig:3Sol_Spectrogram_Last}(b) for z=28cm, where the trailing edge of pulse 2 below 2.3$\mu \mathrm{m}$ (delayed more than 35ps) has now been completely swallowed by pulse 3. The cut-off wavelength for energy transfer from pulse 2 (pulse 1) to pulse 3 (pulse 2) has now increased to 3.0$\mu \mathrm{m}$ (2.8$\mu \mathrm{m}$), as marked by the 10THz Raman indicators.   %
In Fig. \ref{fig:3Sol_Spectrogram_Last}(c) for z=36cm pulse 2 has now been completely swallowed by pulse 3 at wavelength shorter 2.8$\mu$m (delayed more than about 18ps). The Raman indicator between pulse 1 and 3 further shows that pulse 3 now also can swallow energy directly from pulse 1 at wavelengths below 2.6$\mu \mathrm{m}$.
In Figs. \ref{fig:3Sol_Spectrogram_Last}(d-e) for z=54cm and 70cm the most delayed pulse 3 is swallowing more and more energy from the other two and at 70cm both pulse 1 and 2 have disappeared at wavelengths below 3.5$\mu \mathrm{m}$, leaving pulse 3 almost alone as a high energy SPM rogue wave  containing more than 89\% of the total energy, where it initially contained 47\% of the energy. No further noticeable energy transfer is expected because of the weak dispersion close to the ZDW, which will prevent further temporal overlap between the pulses. \\
\begin{figure}[htbp]
    \centering
    \fbox{
    \includegraphics[width = 0.96 \linewidth]{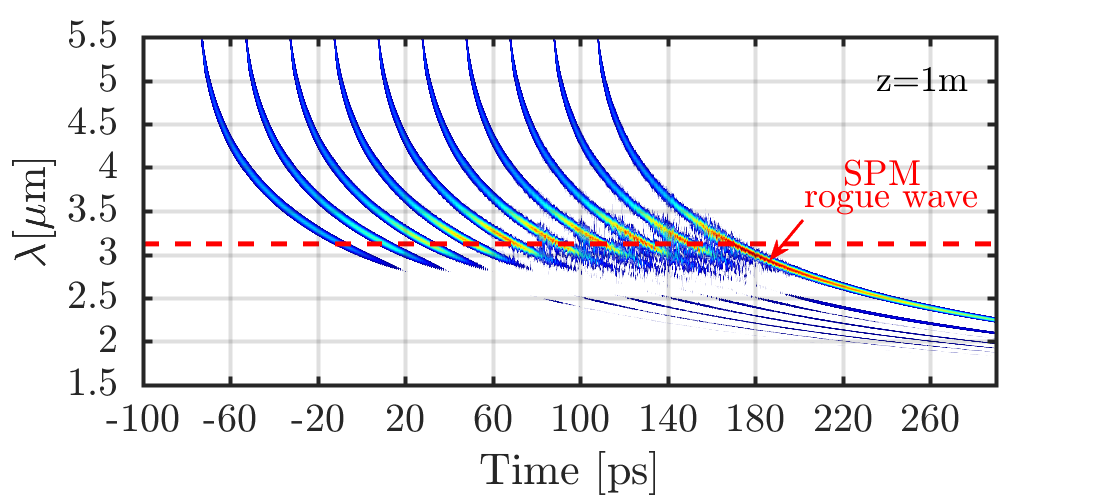}}
    \fbox{\includegraphics[width = 0.96 \linewidth]{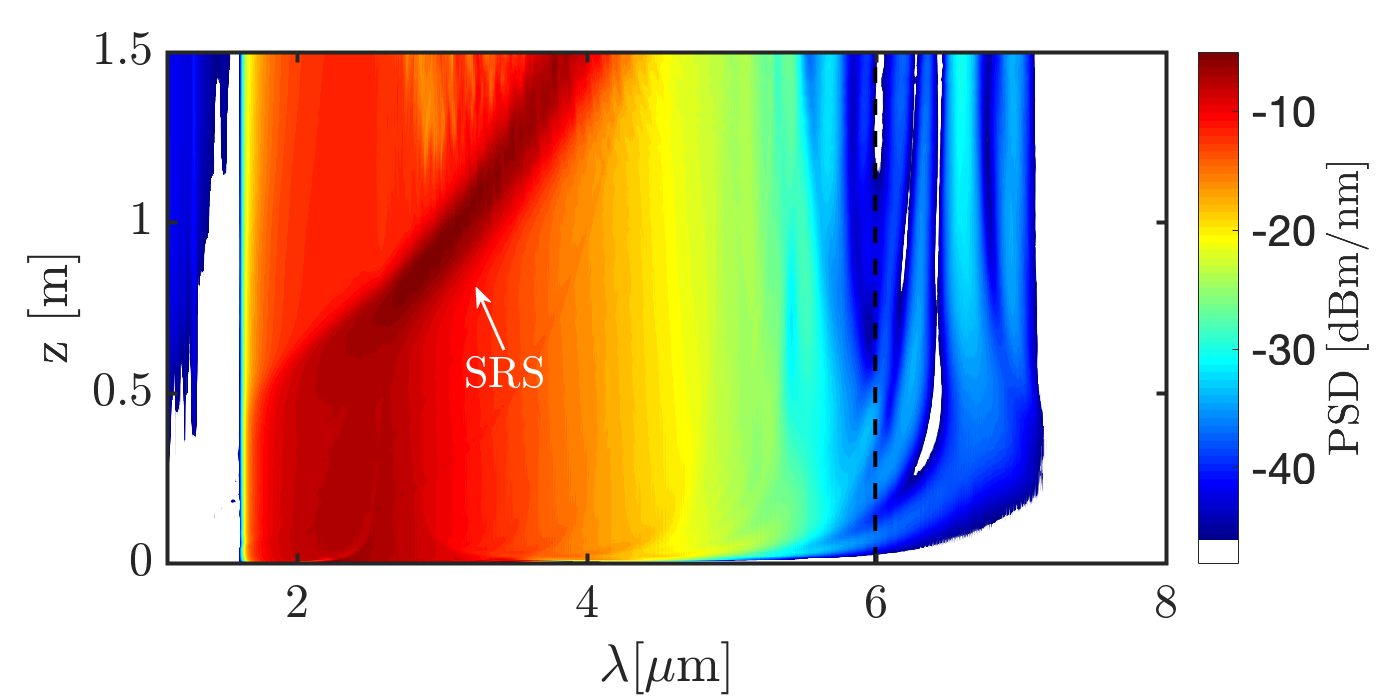}
    }
    \caption{Bottom: Spectral evolution of 10 copies of soliton 3 at 2.68$\mu$m initially separated by 20ps over 1.5m in the $\mathrm{As_2Se_3}$ fiber without loss. Top: Spectrogram at z=1m. The red line indicates the maximum of the PSD. More than 55\% of the total energy is in the trailing pulse at this point.}
    \label{fig:10Solitons}
\end{figure}
The SPM rogue wave is clearly being generated through inter-pulse Raman amplification. To demonstrate the dominant role of SRS we turn off the Raman effect, by setting $f_R=0$, and repeat the simulation out to z=70cm. The result is shown in Fig. \ref{fig:3Sol_Spectrogram_Last}(f), for a direct comparison with Fig. \ref{fig:3Sol_Spectrogram_Last}(e) with the Raman effect present. Without the Raman effect the three SPM pulses are seen to not transfer energy between each other, but be completely intact despite a strong temporal overlap and no SPM rogue wave is generated. The full spectral evolution over 1.5m with the Raman effect turned of is given in the supplementary Fig. S5, which confirms the absence of a collective localized SRC structure when the Raman effect is absent.\\
When the Raman effect is turned off by setting $f_R$=0 obviously both inter-pulse and intra-pulse SRS are turned off. In the supplementary Fig. S6 we therefore show the spectral evolution of a single soliton 3 over 0.7m with the Raman term present and in the available movie we show the evolution of the spectral-temporal structure as spectrograms. Both clearly demonstrate no noticeable intra-pulse SRS, confirming that the high energy SPM rogue wave is generated by inter-pulse Raman amplification.\\
The demonstration of the generation of the high energy SPM rogue wave was performed with 3 different solitons 1-3 from an actual SC appearing in actual realized cascaded mid-IR SCG, in which the rogue wave started as the most delayed soliton, which in this case also had the highest energy already at the input. To even more beautifully demonstrate the generation we simulated the evolution of 10 identical copies of soliton 3 separated by 20ps and set the fiber loss to zero in order to avoid spectral signatures originating for example from loss peaks. The spectral evolution over 1.5m in the lossless $\mathrm{As_2Se_3}$ fiber shown in Fig. \ref{fig:10Solitons} now even more clearly demonstrates the excitation and slow red-shift of the localized SRS structure. More importantly, the SPM rogue wave is still generated as seen in the spectrogram at 1m in Fig. \ref{fig:10Solitons}, where it has swallowed most of the energy of the other 9 pulses and now contains over 55\% of the total energy, where it initially had only 10\% of the energy. The full spectrogram series of this lossless 10-soliton case is shown in Figs. S8-S10 in the supplementary material.

\section{Manipulating the localized SRS structure}
\begin{figure}[htbp]
    \centering
    \fbox{\includegraphics[width = 0.96 \linewidth]{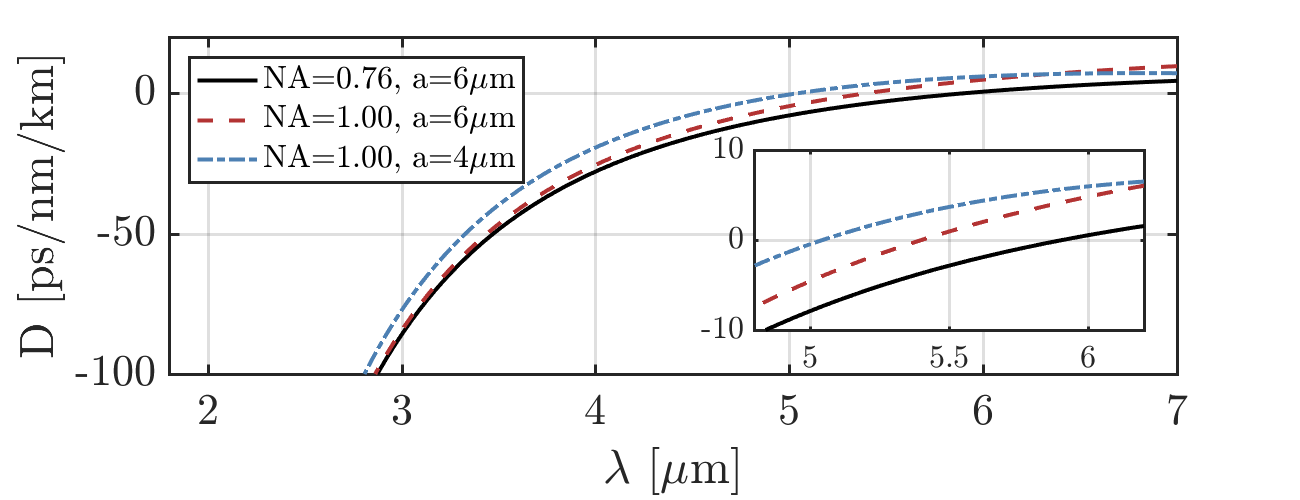}}
    \fbox{\includegraphics[width = 0.96 \linewidth]{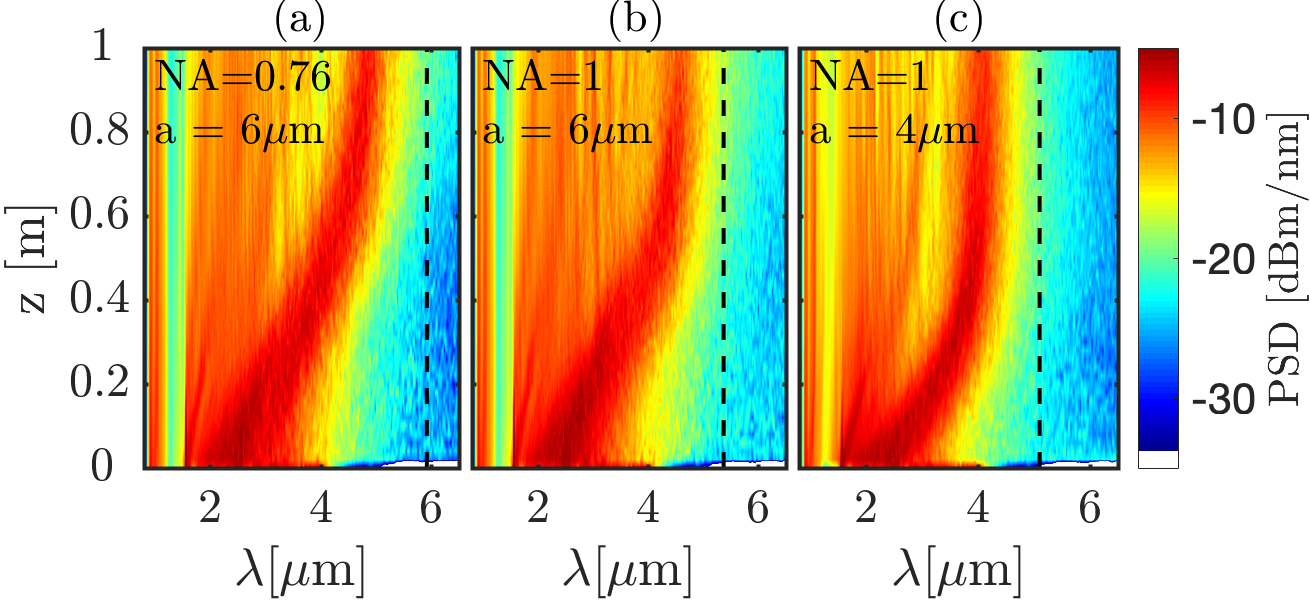}}
    \caption{Bottom: Comparison of the dynamics in 3 $\mathrm{As_2Se_3}$ fibers with different ZDW of approximately 6$\mu$m (original fiber, NA=0.76, core radius=a=6$\mu$m), 5.5$\mu$m (NA increased to 1 by reduced cladding index), and 5.1$\mu$m (NA=1 and reduced a=4$\mu$m), respectively. Top: Dispersion of the 3 fibers with zoom around the ZDWs to see the slope.}
    \label{fig:Parallel}
\end{figure}
Our modelling has clearly shown that the red-shifting localized SRS structure is due to inter-pulse SRS, by which energy is transferred  between SPM pulses towards the most delayed pulse (the SPM rogue wave), and that it is mediated by dispersion continuously causing spectrally longer and longer wavelength parts of the trailing edge of an SPM lobe to begin to overlap with the neighbouring slightly more delayed and red-shifted SPM lobe. This energy transfer between the SPM lobes is also what creates the localized SRS structure and makes it red-shift. If the dispersion is very weak there will be no new energy transfer, since all spectral components travel at the same speed. \\
We thus anticipate that the red-shift of the localized SRS stucture will stop close to the ZDW when the dispersion becomes too weak, and not be able to cross the ZDW. We further anticipate that the bigger the slope of the dispersion at the ZDW ($\beta_3$) is the closer to the ZDW the localized SRS structure can come and the narrower it will be spectrally when it stops. This is exactly what we observe in Fig. \ref{fig:Parallel}, where we propagate the output spectrum of the ZBLAN fiber shown in Fig. \ref{fig:ZBLAN_output}(a) in 1m of the original $\mathrm{As_2Se_3}$ fiber (a) with ZDW=6.0$\mu$m and in 1m of two other $\mathrm{As_2Se_3}$ fibers with increasingly smaller ZDW of (b) 5.5$\mu$m (With increased NA=1 by decreasing the cladding index) and (c) 5.1$\mu$m (NA=1 and decreased core radius a=4$\mu$m). The localized structure never crosses the ZDW. In addition the localized structure comes closest to the ZDW in fiber (b) with the largest dispersion slope at the ZDW of $\beta_3$=2.4ps$^3$/km, where fibers (a) and (c) have $\beta_3$=2.1ps$^3$/km and $\beta_3$=1.9ps$^3$/km, respectively.\\
The ZDW and the slope of the dispersion at the ZDW can thus be used to control and manipulate the SPM rogue wave generation process and the center wavelength and bandwidth of the high PSD localized SRC structure. For example, if a strong signal is required at $7 \mu \mathrm{m}$ for an imaging system similar to the one reported in \cite{Tissue_Imaging}, it would be beneficial to increase the ZDW to approximately $8 \mu \mathrm{m}$, by for example choosing a fiber with $a = 6 \mu \mathrm{m}$, and NA$=0.56$.

\section{Conclusion}

We have through rigorous numerical modelling demonstrated a novel fundamental physical phenomenon – the generation of high energy optical SPM rogue waves in the normal dispersion regime in the form of a strongly SPM broadened pulse containing most of the energy originally spread out over many pulses. Along the local SPM shape the rogue wave is localized both temporally and spectrally, but seen as a whole entity it is both temporally and spectrally delocalized. This is in sharp contrast to the well-known optical rogue waves being generated in the anomalous dispersion regime where solitons and MI exist, taking either the form of the temporally localized high peak power fundamental bright solitons or the both spatially and temporally localized Peregine soliton.
We have demonstrated that this SPM rogue wave is naturally generated from an ensemble of originally separated input pulses that undergo spectral broadening through SPM and temporal broadening through dispersion to temporally overlap and transfer energy between each other through inter-pulse SRS. The inter-pulse SRS transfers energy from one SPM lobe to the neighbouring slightly red-shifted and delayed SPM lobe, so that most of the energy finally becomes localized in the most delayed SPM-shaped pulse - the SPM rogue wave.\\
Technically we have demonstrated the generality of the SPM rogue wave by generating it from different ensembles of pulses and, in particular, by showing how it is naturally appearing in fiber-based cascaded SCG in which the distributed soliton spectrum of an SC generated in the anomalous dispersion regime of one fiber is coupled into the normal dispersion regime of a subsequent fiber. Cascaded SCG is the most promising technique behind the future table-top and low-cost mid-IR
fiber-based SC sources, which are spatially coherent and have a brightness two orders of magnitude higher than synchrotrons \cite{Syncrotron}. Understanding and being able to control and manipulate the SPM rogue wave is therefore not just of fundamental but also advanced technical importance. We have shown how the SPM rogue wave generation can be seen spectrally as a localized collective high PSD structure that slowly red-shifts due to the energy transfer by inter-pulse SRS, until finally being stopped by the ZDW. We have demonstrated how the SPM rogue wave and this localized high PSD structure can be manipulated through the dispersion to tailor the spectrum of a mid-IR SC source towards applications in for example imaging and spectroscopy.

\section*{Funding Information}
We acknowledge the financial support from Innovation Fund Denmark through UVSUPER Grant No. 8090-00060A.

\section*{Acknowledgements}
We would like to thank Kyei Kwarkye for measuring the ZBLAN loss curve, and Mikkel Jensen for fruitful discussions.

\section*{Disclosures}
The authors declare no conflicts of interest.

\section*{Supporting content}
See supplement 1 for supporting content.

\bibliographystyle{unsrt}
\bibliography{main}

\clearpage

\onecolumn
\huge{\textbf{Supplementary document}}
\beginsupplement
\normalsize

\section{fiber parameters}
To allow reproduction we here present the necessary fiber parameters for the entire cascade. In the following we use Si:Er for the initial stage in the fiber cascade, which is an Erbium/Ytterbium doped silica fiber, and Si:Tm for the second stage, which is a Thulium doped silica fiber. The constant fiber parameters for the two fiber amplifiers are given in table \ref{tab:fiber_Param_sup} \\
\begin{table}[htbp]
    \centering
    \begin{tabular}{c c c c c} \hline
     \bettershortstack{fiber \\ \vphantom{$\mu$}} & \bettershortstack{ a \\ $[\mu \mathrm{m}]$} & \bettershortstack{NA \\ \vphantom{$\mu$}} & \bettershortstack{$n_2 \cdot 10^{20}$ \\ $[\mathrm{m^2/W}]$} &
     \bettershortstack{$f_R$ \\  \vphantom{$\mu$}} \\  \hline
    Si:Er     & 6 & 0.2 & 3.2 & 0.18 \\
    Si:Tm    & 5 & 0.22 & 4.23 & 0.13  \\ \hline
    \end{tabular}
    \caption{Constant fiber parameters for the silica gain fibers. where a is the fiber core radius and NA is the numerical aperture. In a SEM analysis conducted in \protect\citeS{RasmusMSC_sup}, it was found that the core of the Thulium doped silica fiber amplifier contains large amounts of Ge. This is the reason why $n_2$ and $f_\mathrm{R}$ differ between the two fiber amplifiers.}
    \label{tab:fiber_Param_sup}
\end{table}
The effective index can be derived from the fiber parameters given in Table 1 of the manuscript, using either analytical results for step-index fibers, or numerical tools, like FEM analysis. The dispersion of both the ZBLAN and the $\mathrm{As_2Se_3}$ fibers are shown in Figure 2, in the main document. \\
The effective area can be derived using the same tools. However, as the effective areas used in the simulations are not included in the main document, we will show them here.

\begin{figure}[htbp]
    \centering
    \fbox{\includegraphics[width=0.8 \linewidth]{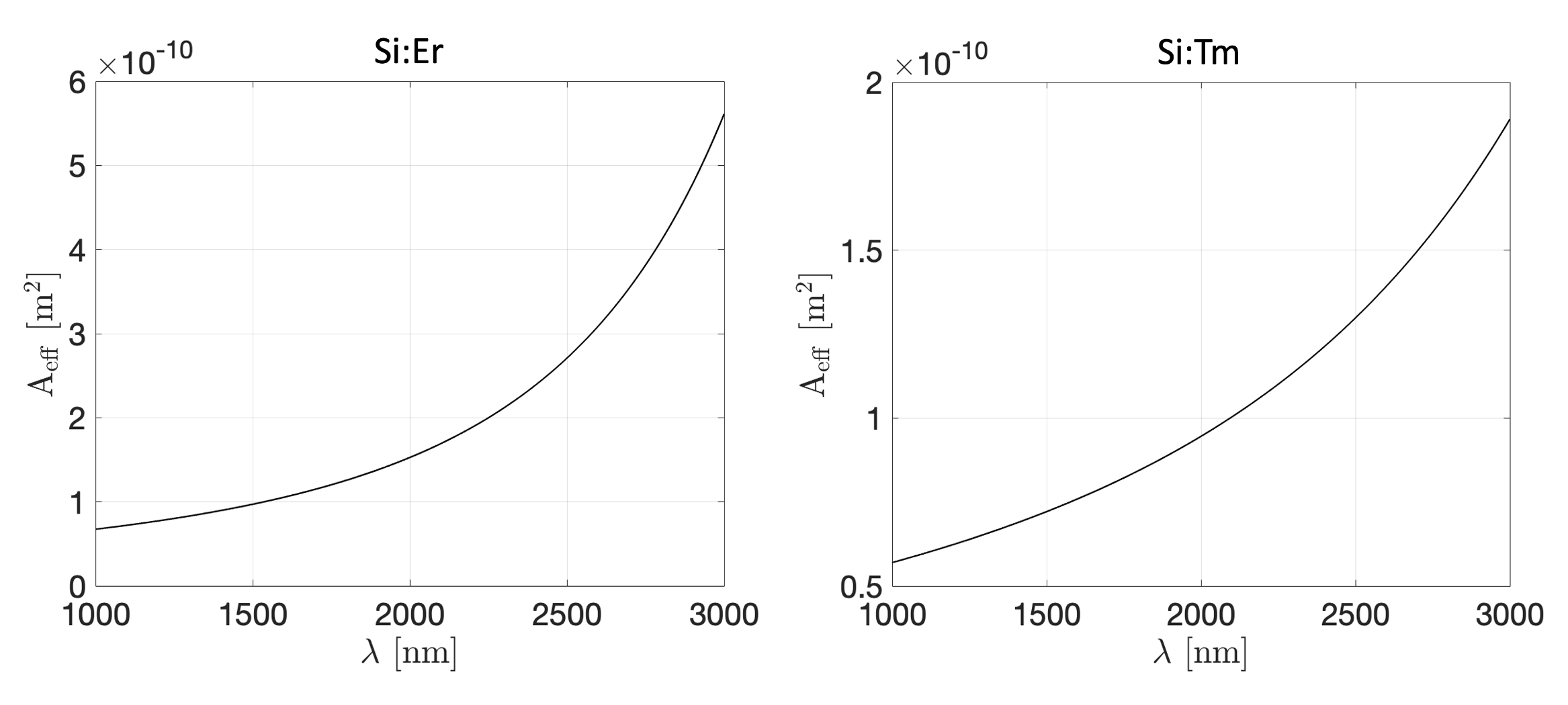}}
    \caption{Left: Effective area of the Si:Er fiber. Right: Effective area of the Si:Tm fiber. These figures are also presented in \protect\citeS{RasmusMSC_sup}.}
    \label{fig:Effective_Area_Si}
\end{figure}

\begin{figure}[htbp]
    \centering
    \fbox{\includegraphics[width=0.4\linewidth]{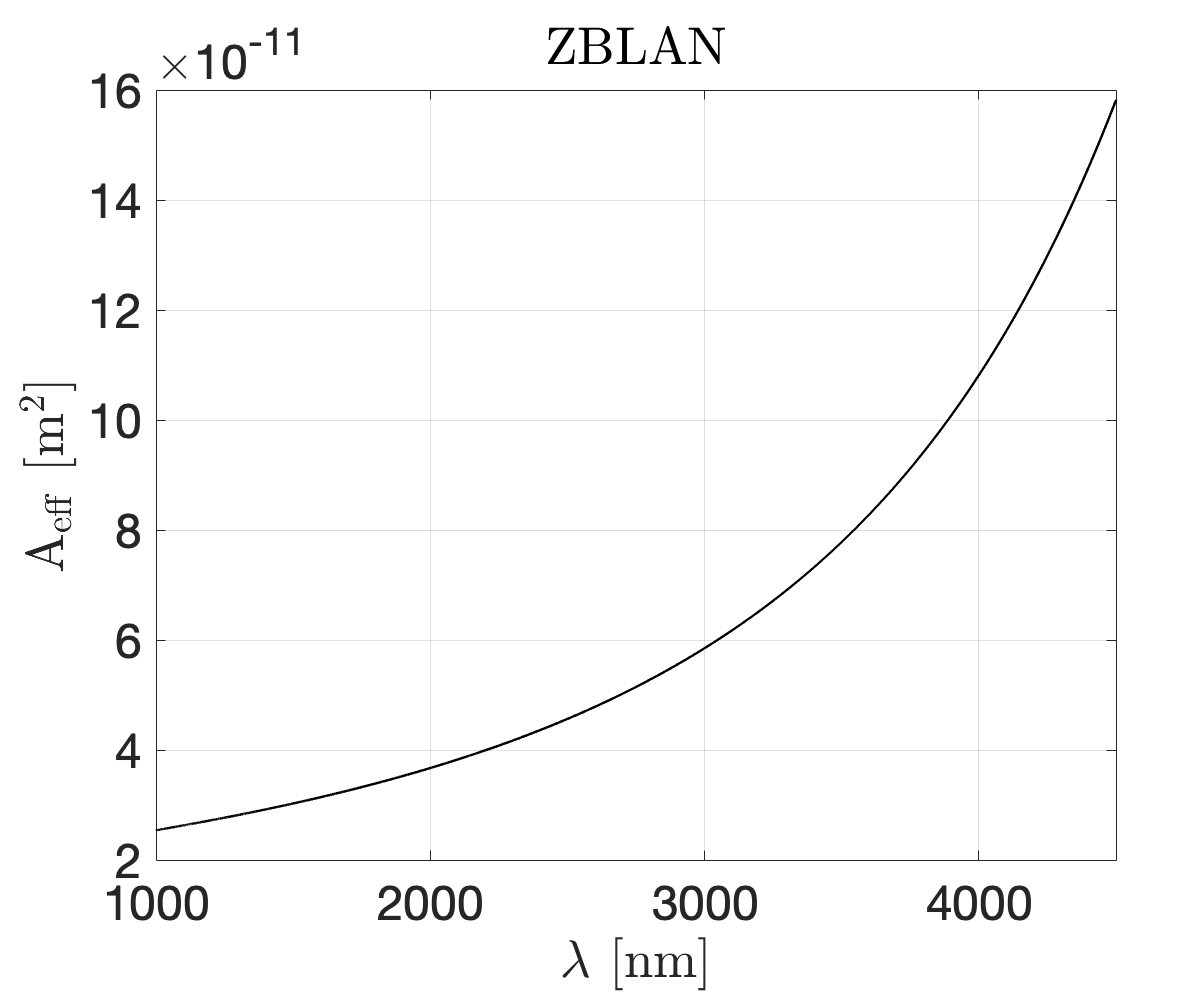}}
        \fbox{\includegraphics[width=0.4\linewidth]{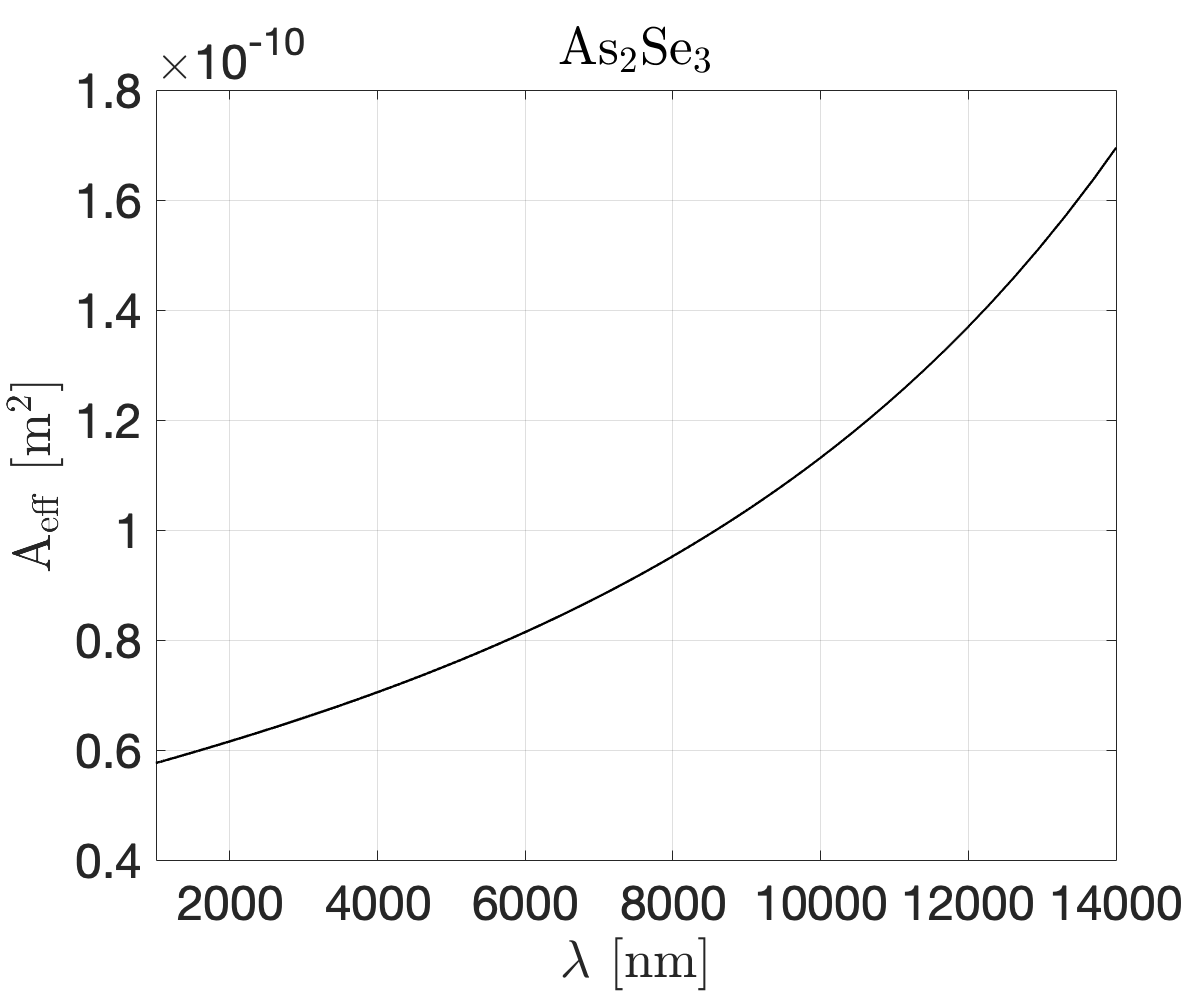}}
    \caption{Left: Effective area of the ZBLAN fiber. Right: Effective area of the $\mathrm{As_2Se_3}$ fiber. This figure is also presented in \protect\citeS{RasmusMSC_sup}.}
    \label{fig:Effective_Area_ZBLAN}
\end{figure}

A parametrisation of the silica loss curve is given in \citeS{MoselundPhd} for both of the fiber amplifiers. The parametrisation is given by:
\begin{equation}
\begin{split}
    &\alpha_{dB/m}(\lambda) = \frac{A_{Ray}}{\lambda^4} + A_{ip} + A_{uv} \exp\qty(\frac{\lambda_{uv}}{\lambda}) + A_{ir} \exp\qty(\frac{-\lambda_{ir}}{\lambda}) \\
    &\alpha_{1/m} = \frac{\ln(10)}{10} \alpha_{dB/m}
\end{split}
\label{eq:si_loss}
\end{equation}
Where
\begin{equation}
\begin{split}
    &A_{Ray} = 1.3 \times 10^{-3} \frac{\mathrm{dB} \mu \mathrm{m^4}}{\mathrm{m}} \quad A_{ip} = 10^{-3} \mathrm{\frac{dB}{m}} \quad A_{uv} = 10^{-6} \mathrm{\frac{dB}{m}} \quad A_{ir} = 6 \times 10^{8} \mathrm{\frac{dB}{m}} \\
    & \lambda_{uv} = 4.67 \mu \mathrm{m} \quad \lambda_{ir} = 47.8 \mu  \mathrm{m}
    \end{split}
\end{equation}

\begin{figure}
    \centering
    \fbox{\includegraphics[width = 0.4 \linewidth]{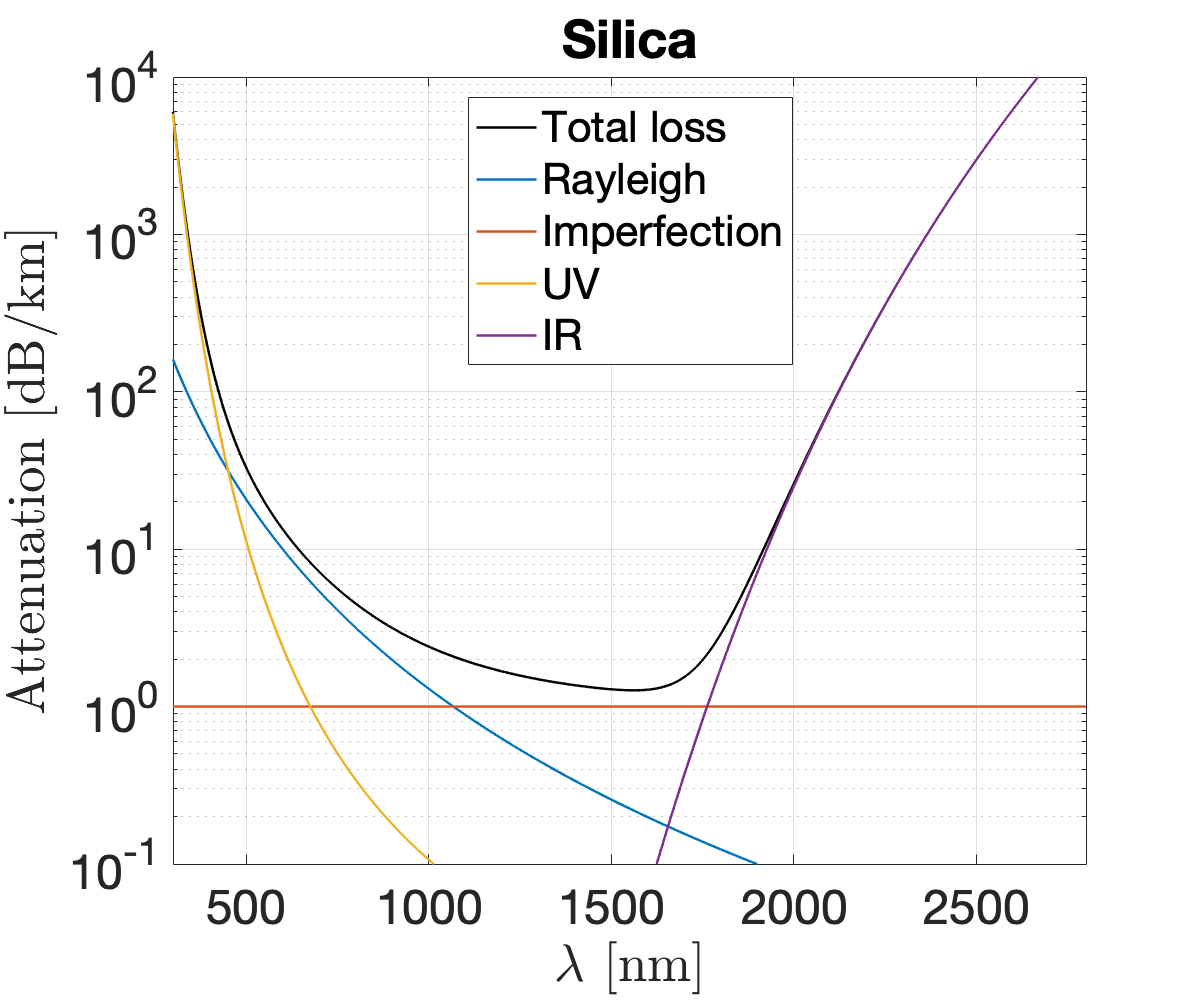}}
    \caption{The used loss curve for the two silica fiber amplifiers. These figures are also presented in \protect\citeS{RasmusMSC_sup}.}
    \label{fig:Si_loss}
\end{figure}

For the Si:Er fiber the gain cross-section is found in \citeS{Digonnet}. The gain cross section for the Si:Tm fiber is given in \citeS{Thulium_Gain}, except that the high wavelength gain is given in \citeS{Digonnet}, the sum of these two curves is the reason for the kink seen around 2100nm.
The gain curves have been rescaled by an average (along the length of the fiber) excited dopant concentration, such that the output spectra of the simulations of the fiber amplifiers matched experimentally measured data as well as possible. As such a constant gain is used in the fibers, which is obviously a hard assumption. 

\begin{figure}[htbp]
    \centering
    \fbox{\includegraphics[width=0.4\linewidth]{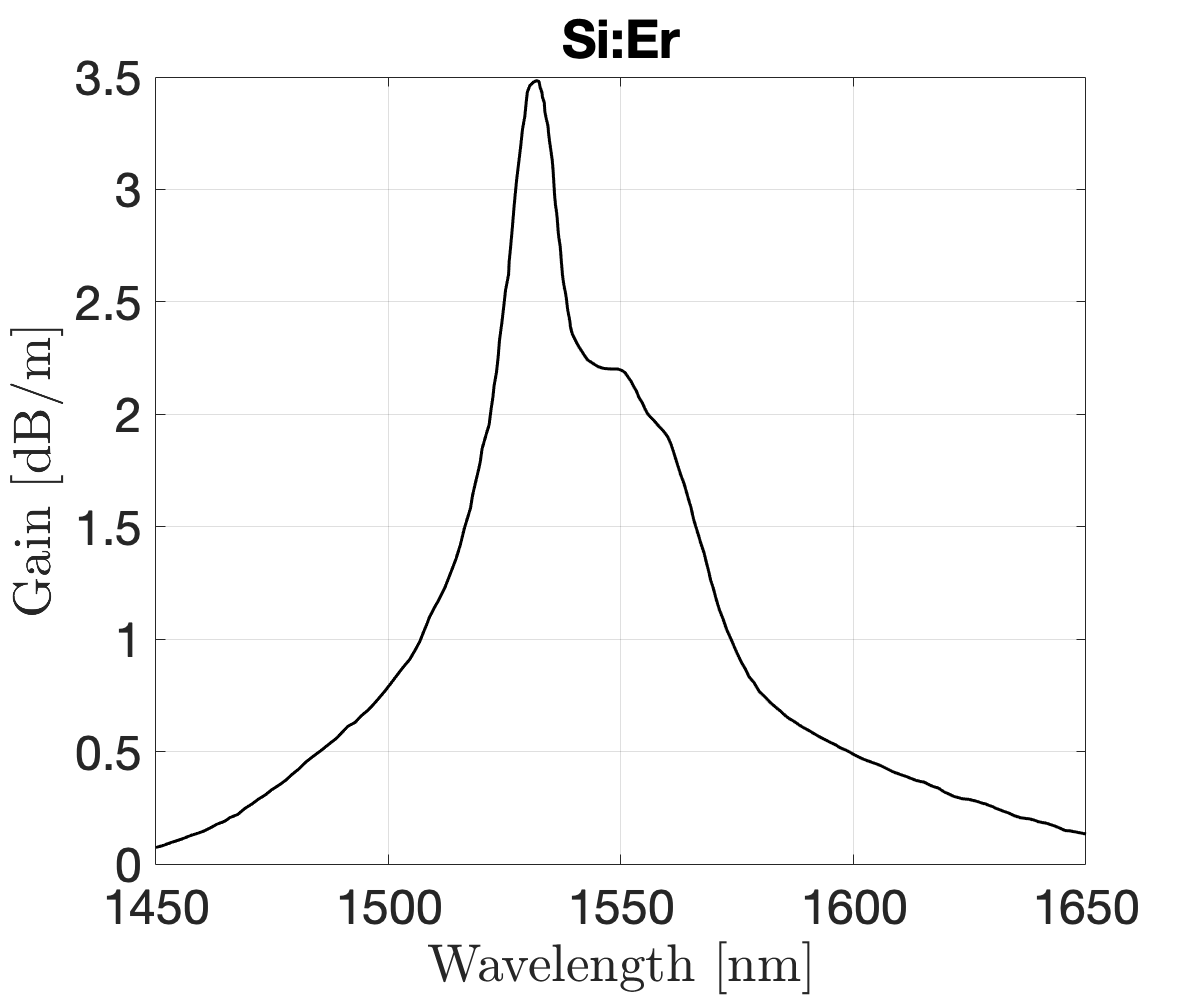}}
        \fbox{\includegraphics[width=0.4\linewidth]{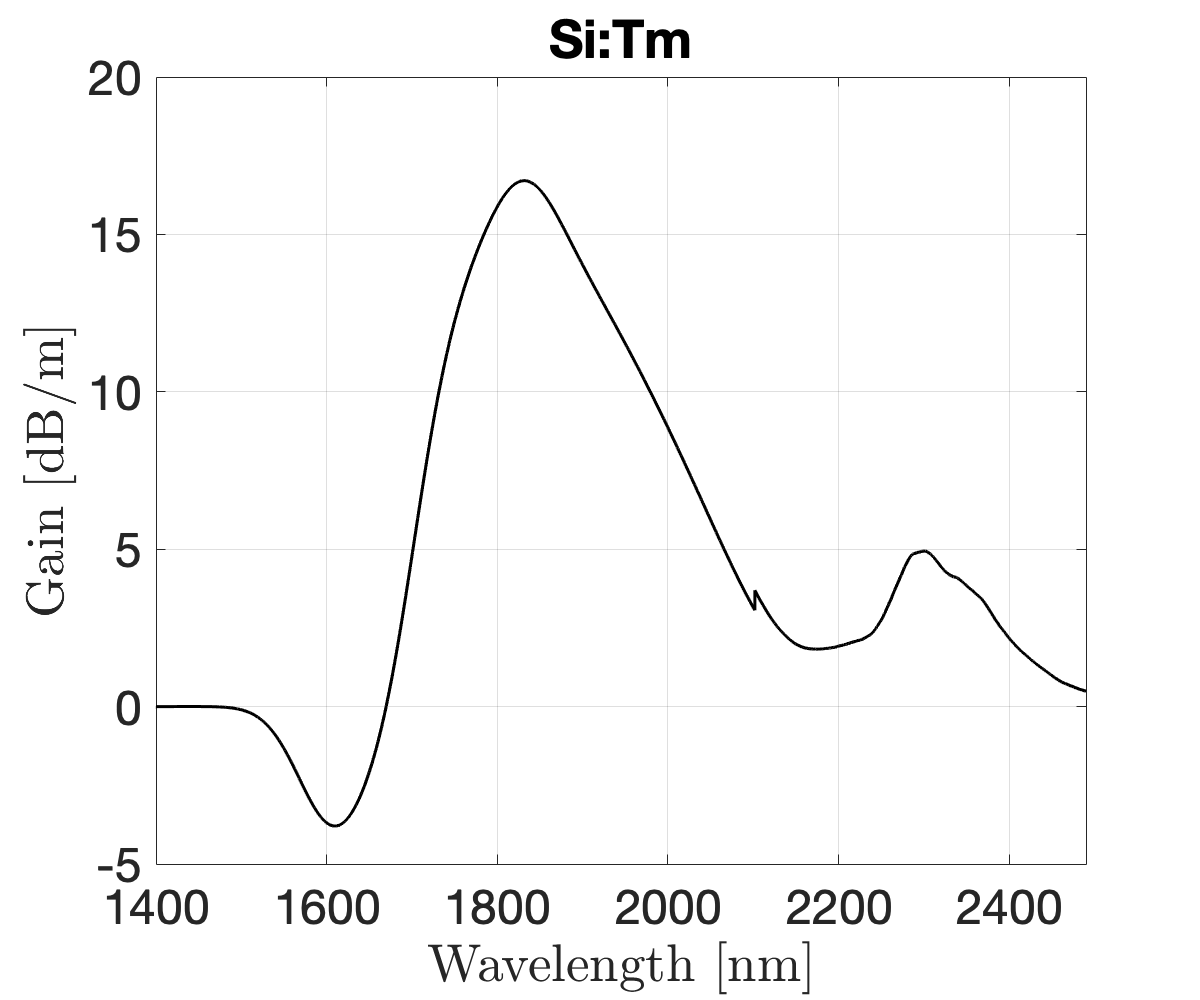}}
    \caption{left: Gain curve for the Si:Er fiber. Right: Gain curve for the Si:Tm fiber. These figures are also presented in \protect\citeS{RasmusMSC_sup}.}
    \label{fig:Gain_Curves}
\end{figure}

 \clearpage

 \section{Three solitons with Raman interaction turned off}
A simulation where the Raman effect was turned off ($f_R = 0$) was run, to clearly show which effects are due to Raman, and which are purely Kerr interaction.

\begin{figure}[htpb]
    \centering
    \includegraphics[width = 0.7 \linewidth]{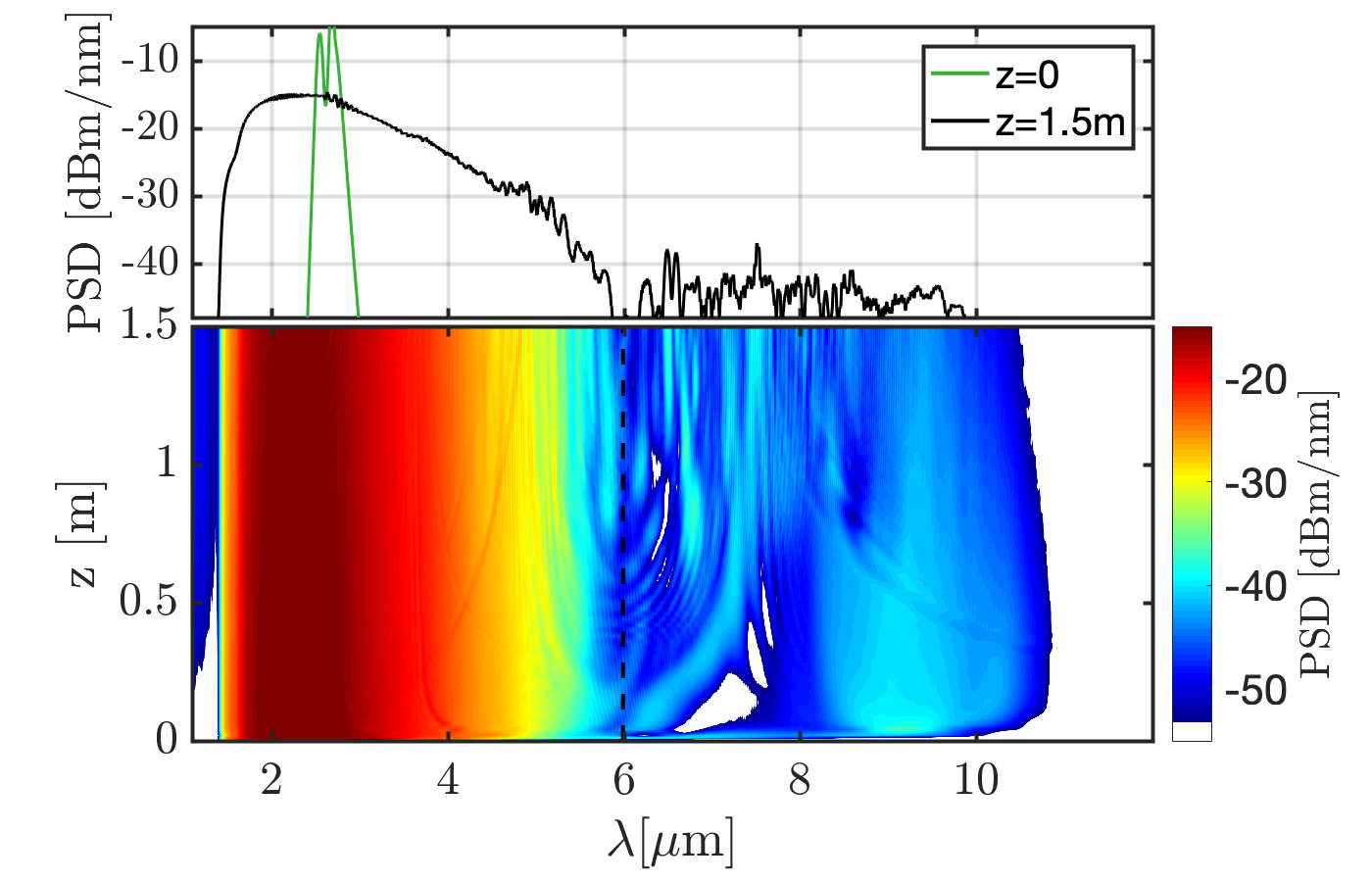}
    \caption{The spectral dynamics of the three test solitons, with the Raman effect turned off ($f_R= 0$). The dynamics are similar to the dynamics in figure 6, except that the SRS structure, seen in figure 6, is lacking. This further confirms that the Raman effect is the cause of the red-shift.}
    \label{fig:No_Raman}
\end{figure}

\clearpage

\section{Simulation with a single input soliton}
\begin{figure}[htbp]
    \centering
    \includegraphics[width=0.7 \linewidth]{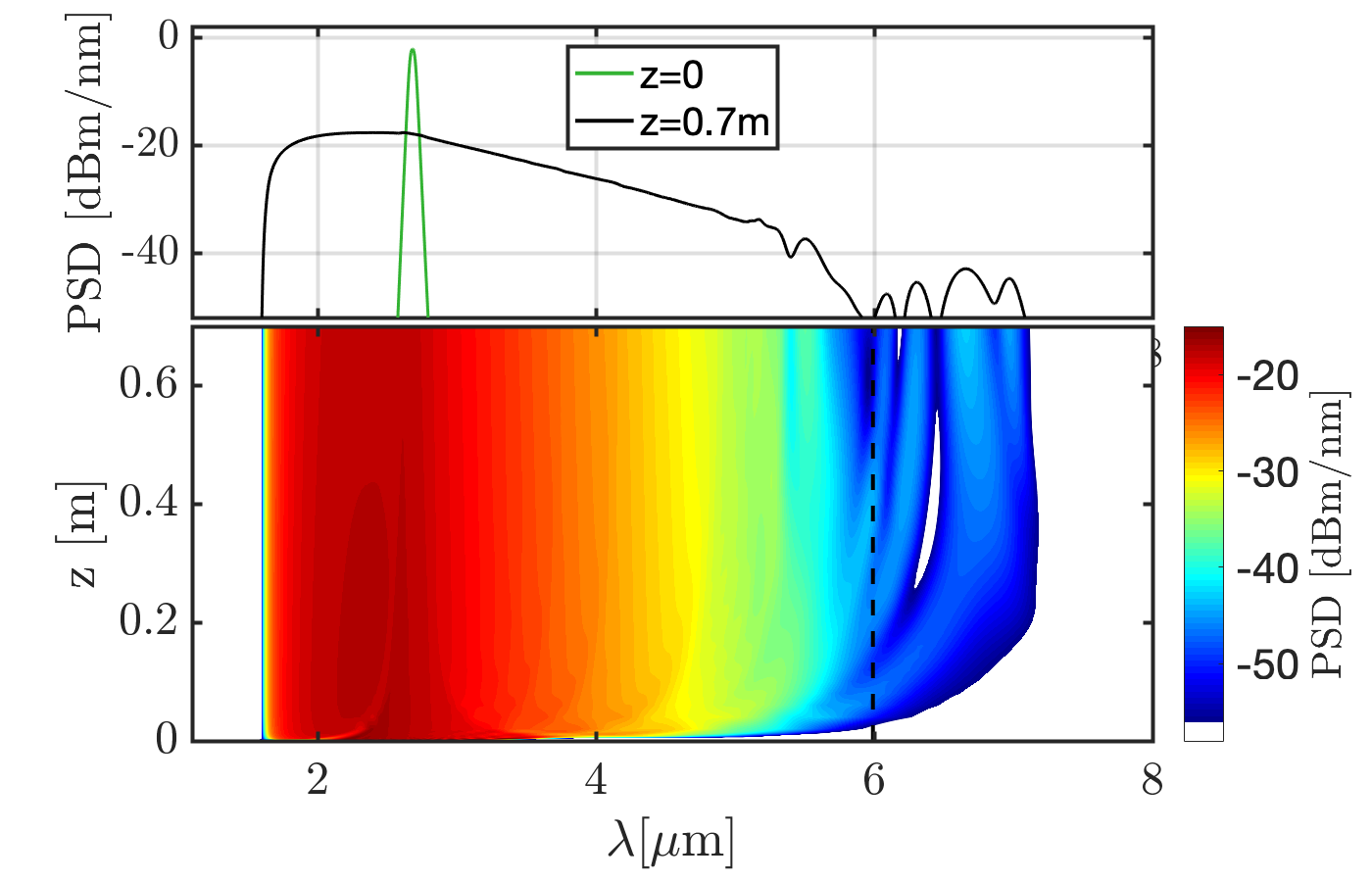}
    \caption{Bottom) The spectral evolution of a single input soliton (soliton "3"). Initially, the dynamics are the same as for 10 input solitons. However, no continuous red-shift is seen in the normal dispersion region. Top) The input spectrum, and the spectrum after 0.7 meters.}
    \label{fig:Single_Sol}
\end{figure}

To underline that the red-shift is indeed a collective effect, we show the dynamics of a single soliton (soliton "3"), in figure \ref{fig:Single_Sol}. \\

\clearpage

\section{Ten similar solitons, with no loss}
\begin{figure}[htbp]
    \centering
    \includegraphics[width = 0.7 \linewidth]{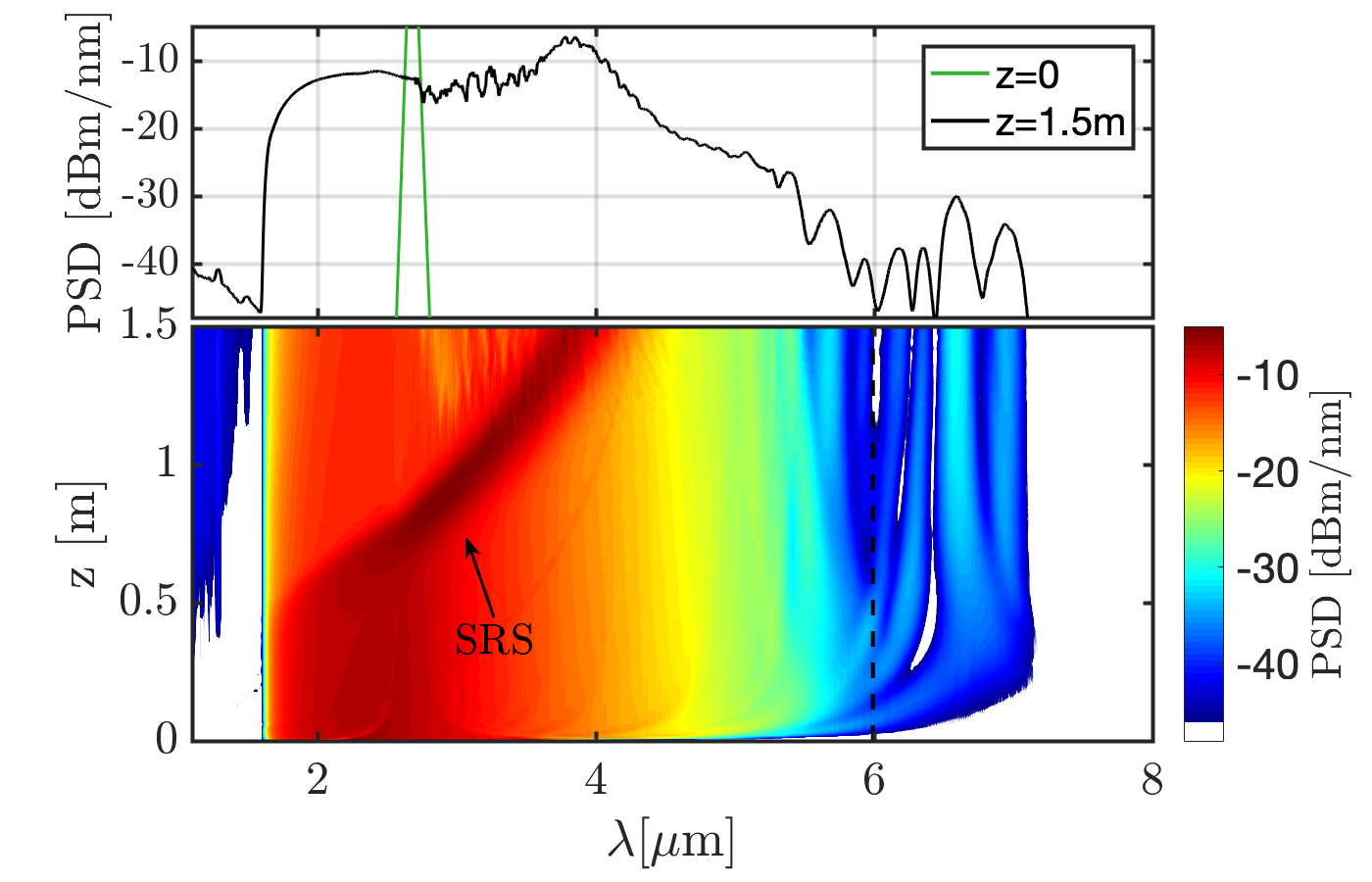}
    \caption{The spectral dynamics of 10 identical solitons (soliton "3"), without loss, and initially temporally shifted by 20ps. The attained spectral width in this case is narrower, as soliton "3" has lower peak power than soliton "2".}
    \label{fig:10Solitons_Dynamics}
\end{figure}

As an additional illustrative example of SRS we  present the dynamics of 10 spectrally identical solitons, at $\lambda_0 = 2.68 \mu \mathrm{m}$ initially shifted by 20ps, in figure \ref{fig:10Solitons_Dynamics}. This is the same simulation as shown in Figure 9 in the main document, but here we show it in more detail. The fiber parameters are all the same as the ones used in the main paper. The spectral dynamics are very similar to the test example of soliton "1,2 and 3", except for that the SRS structure is seen more clearly. In the following we will show spectrograms breaking down both the temporal and spectral dynamics of the 10 solitons. \\

\begin{figure}
    \centering
    \includegraphics[width=0.5\linewidth]{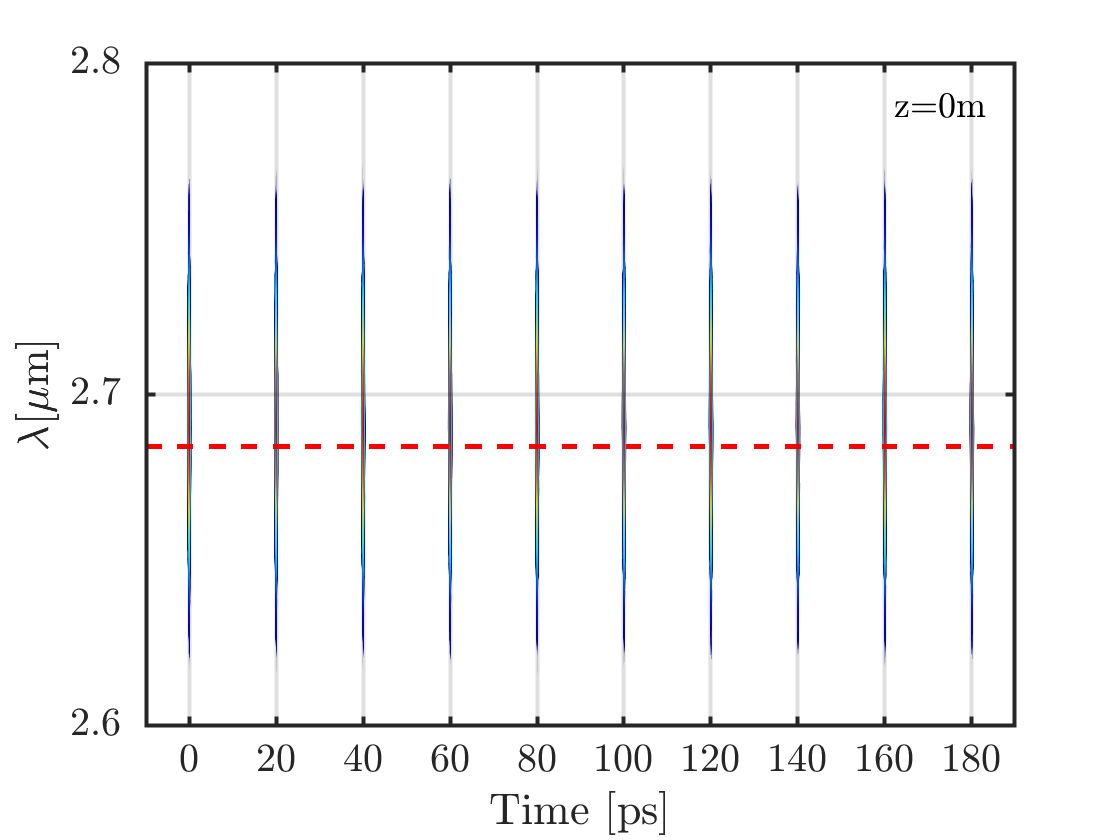}
    \includegraphics[width=0.5\linewidth]{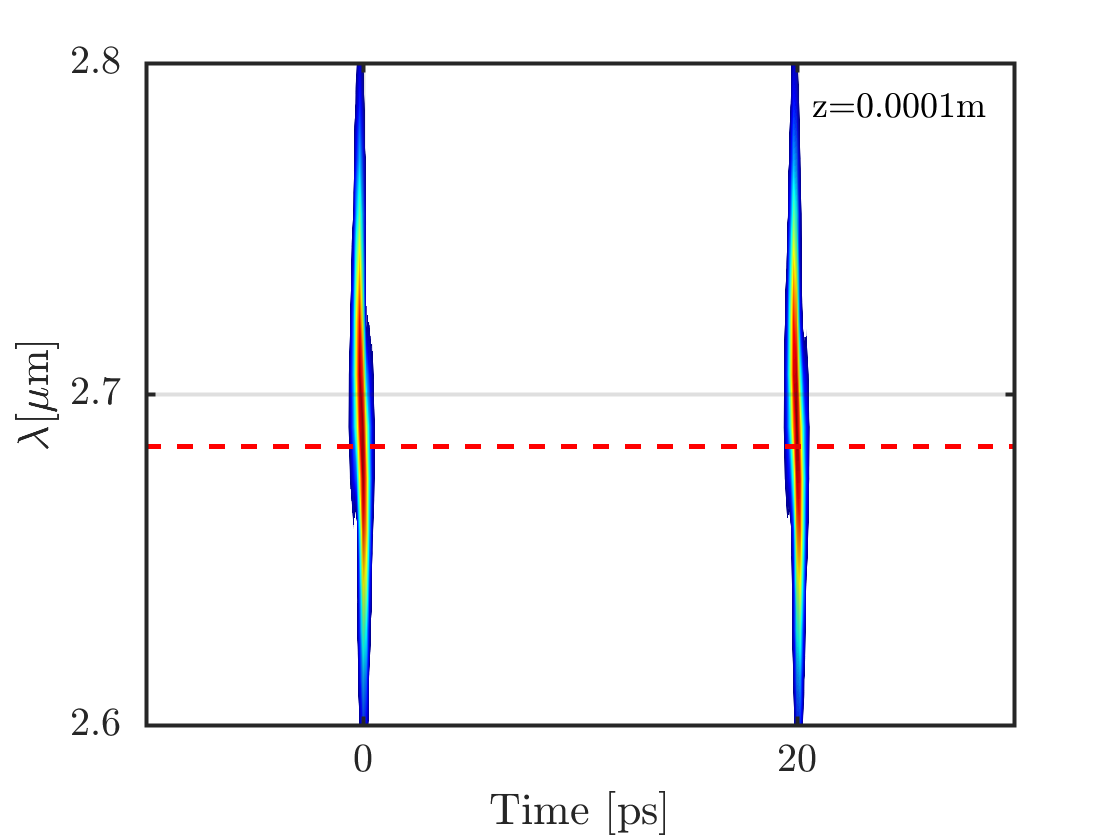}
    \includegraphics[width=0.5\linewidth]{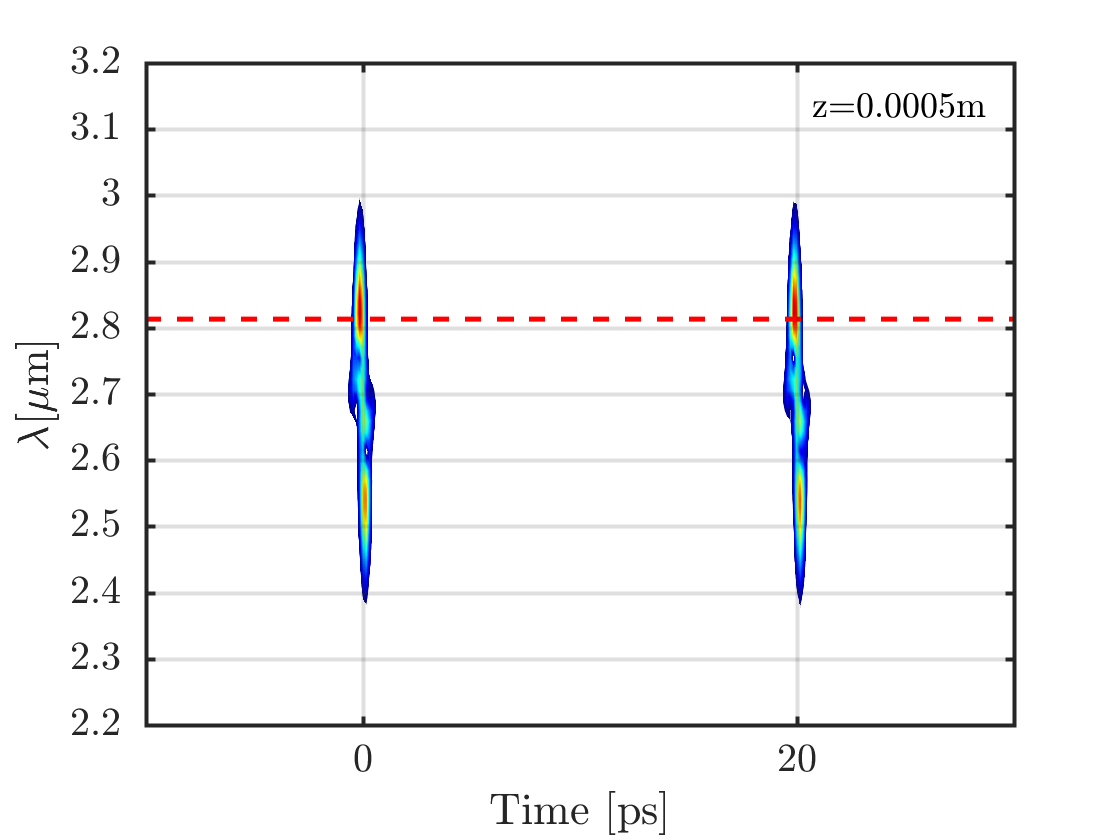}
    \caption{The initial dynamics shown in spectrograms. The red line in the plots is where the maximum PSD is found. In the bottom two plots only two of the pulses are shown in order to clearly illustrate optical wave breaking. In the bottom-most plot the pulse centres are almost depleted, and the main PSD is localised in two new bands.} 
    \label{fig:FirstSpectrograms}
\end{figure}

\begin{figure}
    \centering
    \includegraphics[width=0.5\linewidth]{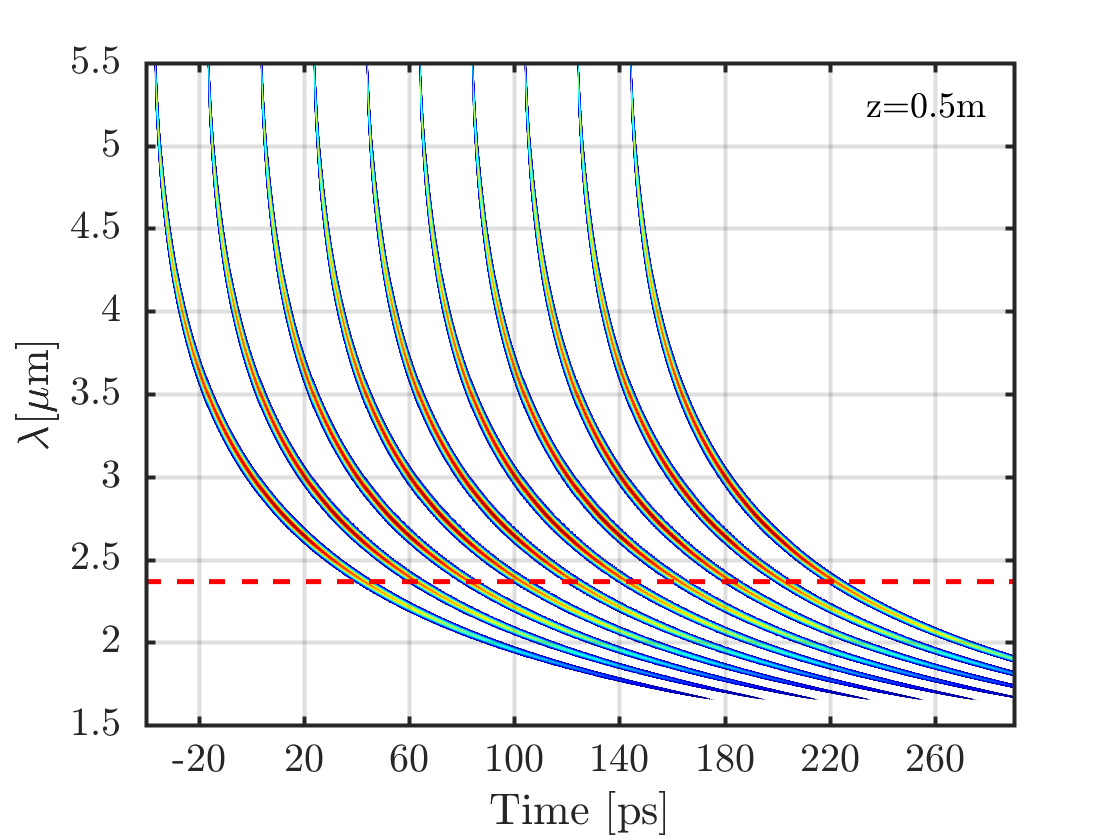}
    \includegraphics[width=0.5\linewidth]{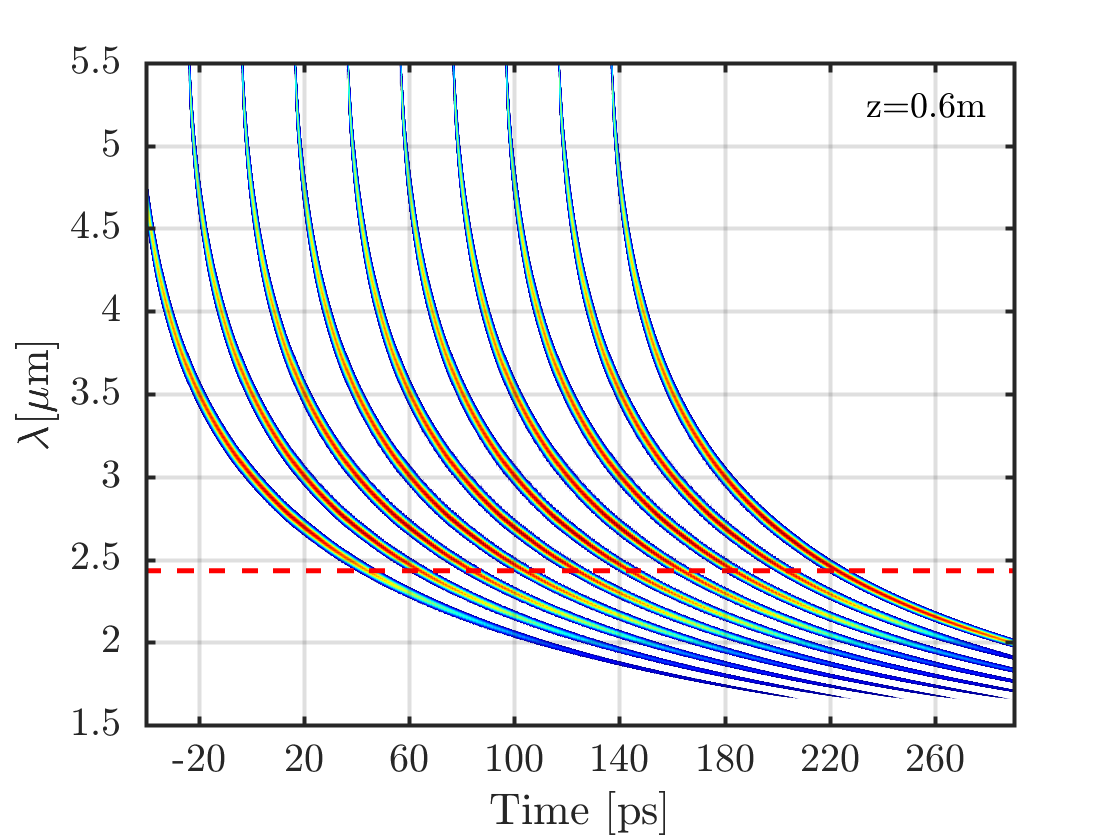}
    \includegraphics[width=0.5\linewidth]{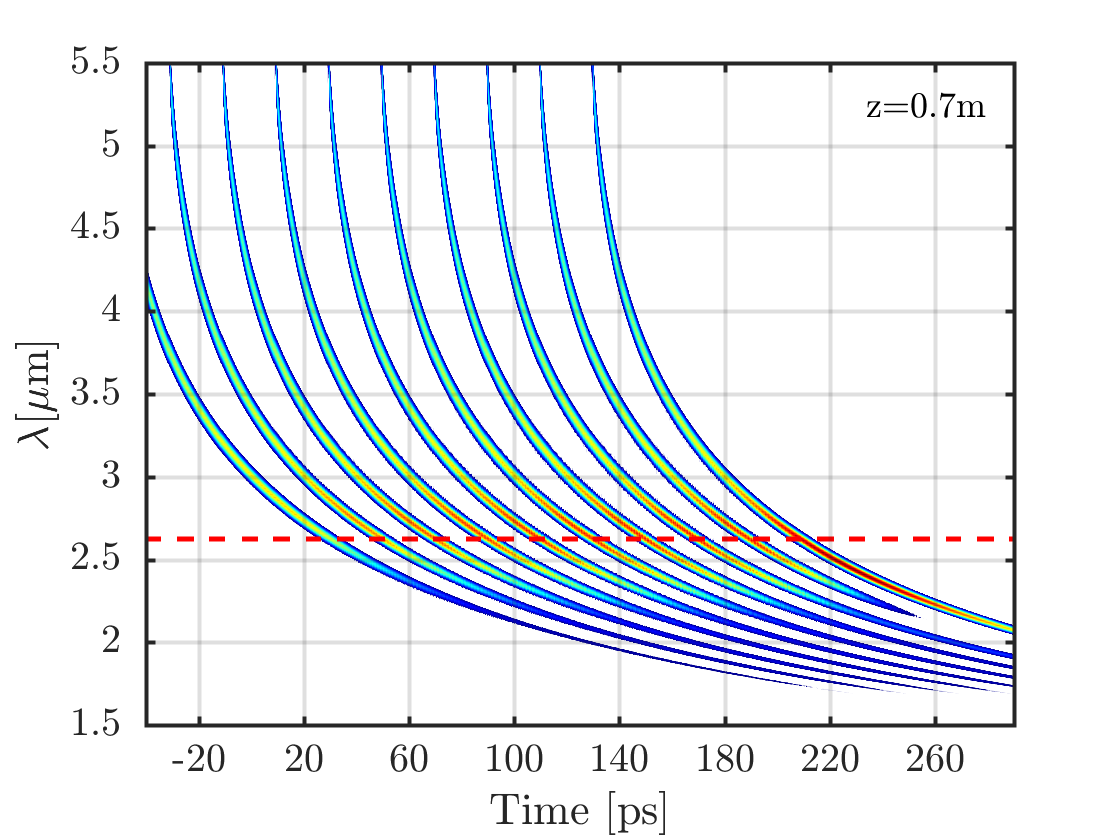}
    \caption{The onset of SRS shown in spectrograms. The red line in the plots is where the maximum PSD is found. At 0.5m the dispersion has not yet ensured the needed overlap for SRS. At 0.6m the red-shift has initiated. This is mostly observable from the depletion of the tail of leading pulse, and from the fact that the maximum PSD is now found at a longer wavelength. At 0.7m the trailing pulse has absorbed most of the short wavelength tail of the neighbouring pulse.}
    \label{fig:MiddleSpectrograms}
\end{figure}

\begin{figure}
    \centering
    \includegraphics[width=0.5\linewidth]{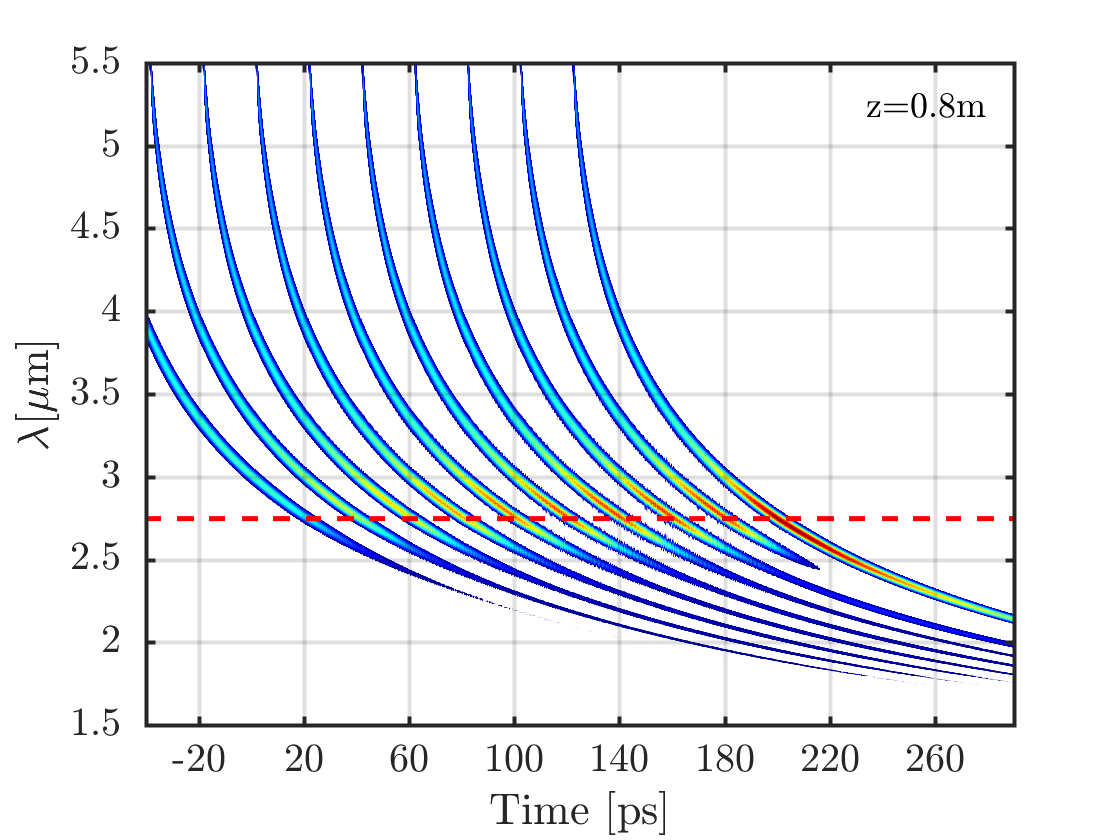}
    \includegraphics[width=0.5\linewidth]{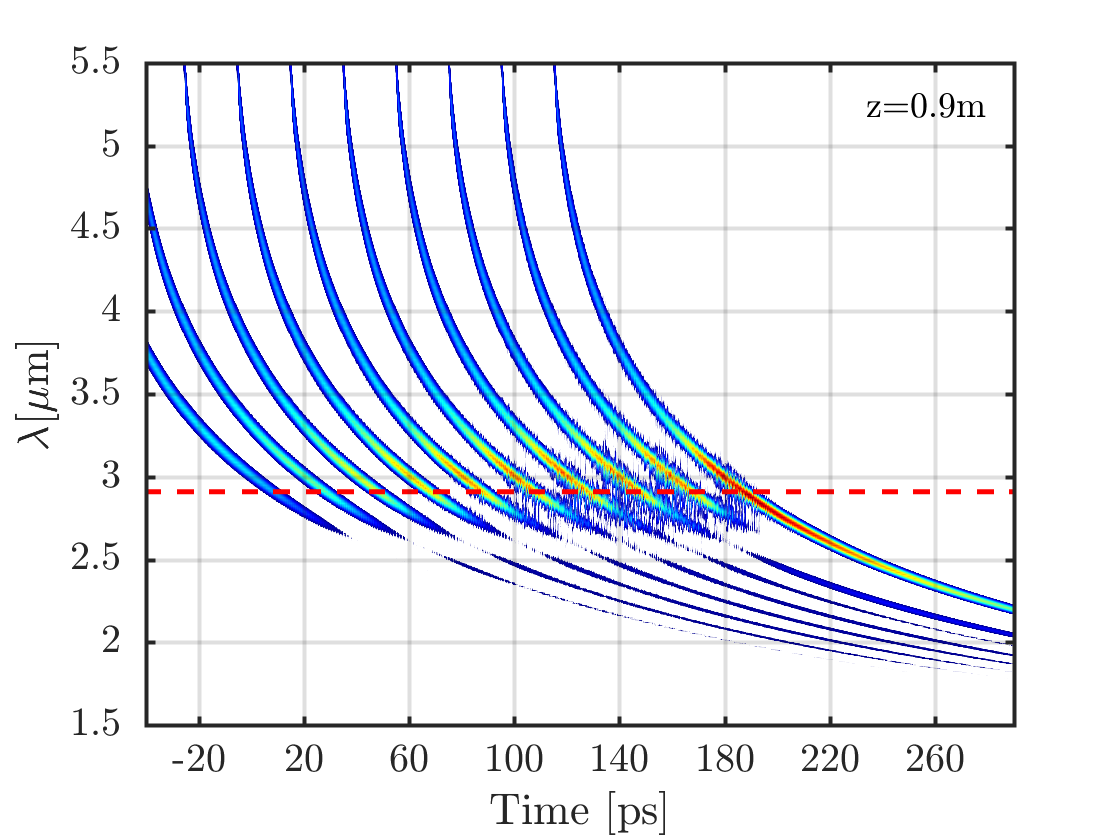}
    \includegraphics[width=0.5\linewidth]{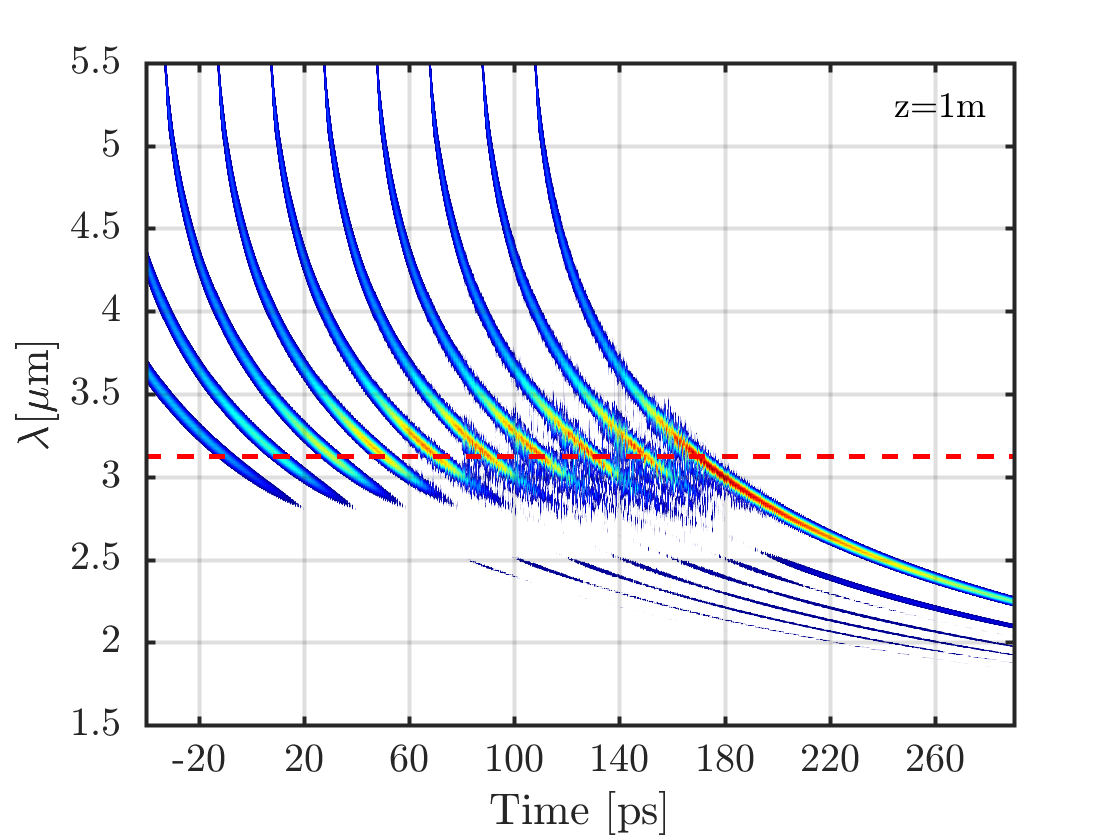}
    \caption{The evolution of SRS shown in spectrograms. The red line in the plots is where the maximum PSD is found. The maximum PSD is continuously shifted towards longer wavelengths. At 0.9m the vacuum noise floor has gained on the pulses, giving a much more noisy spectrogram. This is further enhanced at 1m. It is important to, once again, note that loss was not included in this simulation. As such the extent of the noise is exaggerated.}
    \label{fig:LastSpectrograms}
\end{figure}

\clearpage

\bibliographystyleS{unsrt}
\bibliographyS{main}

\end{document}